\documentclass[%
reprint,
 amsmath,amssymb,
aps,
prb,longbibliography,
floatfix,
]{revtex4-2}
\usepackage{graphicx}
\usepackage[dvipsnames]{xcolor}
\definecolor{myblue}{RGB}{80,150,255}
\usepackage{dcolumn}
\usepackage{bm}
\usepackage{appendix}
\usepackage[hidelinks]{hyperref}
\usepackage{siunitx}
\usepackage{braket}
\usepackage{cancel}

\newcommand{\MIPT}{Center for Photonics and 2D materials, Moscow Institute of Physics and Technology, 141700,  9 Institutskiy per., Dolgoprudny, Moscow Region,  Russia }
\newcommand{\RQC}{Russian Quantum Center, 121205, Bolshoi Boulevard, Building 30,  Skolkovo Innovation Center, Moscow,  Russia}
\newcommand{\ITMO}{Department of Physics and Engineering, ITMO University, 191002, Lomonosova St. 9, St. Petersburg, Russia}
\newcommand{\MIP}{Moscow Center for Advanced Studies, Kulakova str. 20, Moscow, 123592 Russia}
\newcommand{\ind}{Independent researcher, Zelenograd bldg. 701, apt. 81, Moscow 124489, Russia}
\usepackage{amsmath}
\usepackage{import}

\ifdefined\RoundTwoMarkup
  \newcommand{\roundtwo}[1]{\textcolor{red}{#1}}
\else
  \newcommand{\roundtwo}[1]{#1}
\fi
\begin{document}

\preprint{APS/123-QED}

\title{Dipole–exchange spin waves and mode hybridization in magnetic nanoparticles}

\author{
Fedor Shuklin$^{\dagger, 1}$}
\email{fedor.shuklin.95@mail.ru}
\thanks{\\$^1$ F.S. and K.A. contributed equally to this work.}
\author{
Khristina Albitskaya$^{\ddagger,1}$}
\author{
Sergei Solovyov$^{\S}$,
Alexander Chernov$^{\dagger,\parallel}$,
Mihail Petrov$^{\P}$
}

\affiliation{$^{\dagger}$\MIPT}
\affiliation{$^{\ddagger}$\MIP}
\affiliation{$^{\S}$\ind}
\affiliation{$^{\parallel}$\RQC}
\affiliation{$^{\P}$\ITMO}

\date{\today}
\begin{abstract}

We investigate spin-wave modes in confined ferromagnetic resonators with spherical and cylindrical geometries across the exchange-dominated, dipole–exchange, and dipolar interaction regimes. Starting from the linearized Landau–Lifshitz-Gilbert equation, we show that the projection of the total angular momentum and mirror parity are conserved quantities in the problem of axially symmetric resonators. These symmetries provide a natural classification of spin-wave modes and explain the degeneracy of exchange modes, as well as its lifting by dipolar interactions. Numerical analysis shows that the nonlocal dipolar interaction removes the exchange degeneracy and hybridizes modes, leading to avoided crossings between modes that belong to the same symmetry sector. To describe this behavior, we develop a coupled-mode theory formulated directly in terms of dynamical magnetization, which reduces the dipole–exchange problem to a finite system of interacting modes. The resulting framework provides a unified description of spin-wave spectra in confined magnetic particles from the exchange limit to the dipolar regime.

\end{abstract}
\maketitle
\section{Introduction}
\label{sec:intro}
Magnetic structures ranging from nanometer to micrometer scales play a crucial role in modern nanophysics \cite{fernandez2017three}. 
Their applications extend from magnetic resonance imaging \cite{konnova2024magnetic}, biomedical technologies \cite{tiwari2022magnetic}, anticancer therapy, and to magnetic-power-independent memory elements \cite{ishak1988magnetostatic, gertz2014magnonic}, telecommunications \cite{chumak2017magnonic, krawczyk2014review, PhysRevApplied.22.014037}, sensing and metrology \cite{Barulin-me-review, haas2022development}, and continue to be developed to this day.

In microwave magnonics the characteristic linear dimensions of the structures are set by the operating wavelength and typically range from several tens of micrometers to millimeters \cite{gurevich2020magnetization, vugal1989magnetostatic}. These dimensions are much larger than the exchange lengths of the materials commonly used in resonant ferromagnetic systems, so the magnon dynamics is dominated by dipolar interactions \cite{vugal1989magnetostatic, stancil2012theory}.

In recent years, however, the demand for more compact, energy-efficient, and high-performance devices \cite{rezende2020fundamentals} has driven the development of micromagnonics, where the system's dimensions scaled down to the micrometer scale. Moreover, the rapid expansion of optical communication technologies and photonics operating at wavelengths around \(1.5\,\mu\mathrm{m}\) and below has introduced  additional challenges. Merging micromagnonic systems with photonic devices gave a strong momentum to the emergence of the area of optomagnonics \cite{Wang_Gao_Jiao_Wang_2022, PhysRevA.94.033821, Chernov:17}.

\begin{figure}[t!]
    \centering
    \includegraphics[width=1\columnwidth]{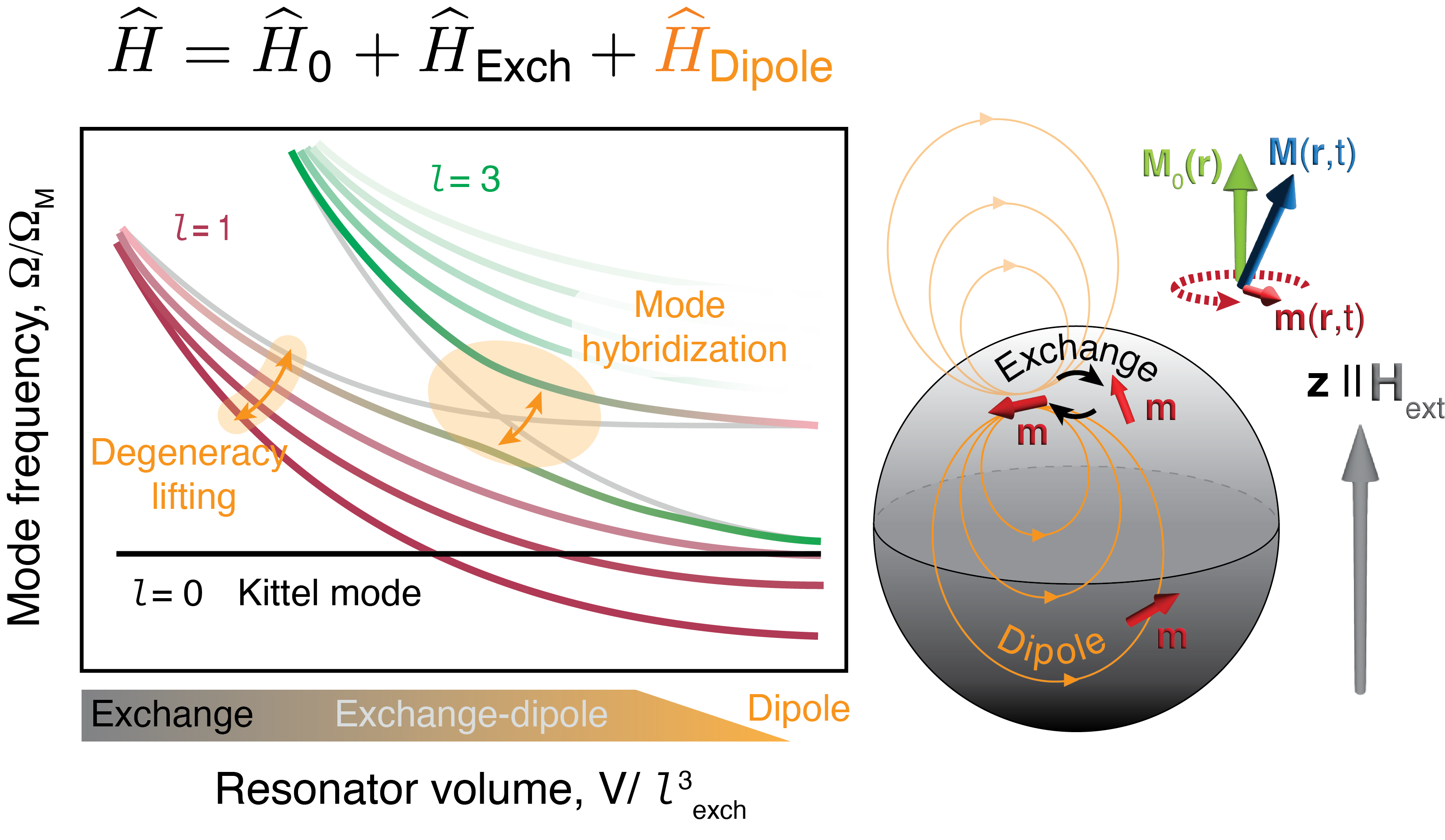}
    \caption{ \label{fig:Main} Schematic evolution of spin-wave modes in a ferromagnetic resonator as the system size increases from the exchange to dipole-dominated regimes. Dipolar interactions lift the exchange degeneracy of modes labeled by $l$ (orbital momentum) and induce hybridization, producing avoided crossings, while the uniform $l=0$ Kittel mode remains unchanged in spherical particle.
     }  
\end{figure}
Optomagnonic interactions are typically enabled by magneto-optical phenomena such as the inverse Faraday effect (for circularly polarized light) \cite{van1965optically} and the inverse Cotton–Mouton effect (for linearly polarized light) \cite{blank2023effective}. Such coupling mechanisms represent a compelling form of light–matter interaction and have been investigated in a variety of plasmonic \cite{Ignatyeva2019-yy,Khramova:19, f6cr-tfld} and dielectric \cite{fan2019magneto} systems like thin films \cite{pantazopoulos2020planar}, waveguides \cite{zhu2022inverse}, or resonators whose spatial dimensions exceed optical wavelengths by several orders of magnitude \cite{Pirro2021AdvancesIC, rezende2020fundamentals}. That indicates significant potential for further miniaturization.
 
Recently proposed compact optomagnonic single Mie scatterers \cite{ignatyeva2023optical,alpakis-sph, alpakis-non-sph, belotelov-spin} and resonant Mie metasurfaces\cite{10.1063/1.5066307,10.1063/5.0257020,ZimnyakovaIgnatyevaKarkiVoronovShaposhnikovBerzhanskyLevyBelotelov, doi:10.1021/acs.nanolett.0c01528} have attracted growing interest as promising candidates for compact optomagnonic devices, since their submicron dimensions are comparable to optical wavelengths.  Optical ferromagnetic Mie scatterers have been theoretically predicted to exhibit magnon-assisted nonlinear photon transitions, magnon–magnon coupling \cite{zhang2026perspective,chen2018strong,sud2020tunable} and nonreciprocal optical responses \cite{PhysRevB.90.174416}. 

In this context, individual resonators with submicrometer dimensions constitute fundamental building blocks of the emerging fields of micro- and optomagnonics and therefore require a precise characterization of their spin-wave mode spectra as well as robust and systematic methods for their theoretical description. 
At these length scales, a comprehensive treatment of spin-wave resonators must explicitly account for the interplay between exchange-dominated and dipole-dominated interaction regimes.
Consequently, a rigorous analysis of spin-wave dynamics in magnetic structures with simple geometries, such as spheres and cylinders, is of fundamental importance. 
Although the underlying problem formulation appears straightforward, spin waves in three-dimensional magnetic submicron resonators remain insufficiently explored, and the field still lacks a detailed mode classification, including a systematic understanding of the role of symmetries, conserved quantities, and the organization of mode families across different interaction regimes.

The groundwork in this direction has started from the seminal papers by Walker \cite{walker1958}, Fletcher, and Bell \cite{fletcher1959}, who developed a theory of dipole-dominated magnonic modes in large spheroids where the magnetostatic approach is fully valid. They have constructed and successfully solved an equation for the scalar magnetic potential. While rigorous solutions beyond spheroid geometries are hard to find,  Yukawa and Abe proposed a semi-analytical theory for finite ferromagnetic cylinders,  calculating magnonic modes by numerically solving the integral equation for magnetization deviation \cite{yukawa1974fmr}. 

The opposite approach considers the systems sufficiently small that the contribution of the exchange interaction dominates over the dipolar one, thus entering the exchange-dominated regime. For ferromagnetic spheres, the first analytical ansatz was proposed  by Amikam Aharoni \cite{aharoni1991exchange}, but focusing only on angular-independent modes with orbital number $m=0$. Aharoni's approach was later generalized in the works of Mills, Chu, and Arias, who extended the analysis to all types of spin-wave modes in a sphere. Building on these results, Mills and Arias studied exchange and dipolar contributions to magnonic spectra of spheres and the dependence of modes' properties on various material parameters \cite{arias2005dipole}.  
While these studies have  addressed the issue of exchange-dipole competition in spherical geometry, their approach lacks a clear picture of the transition from the exchange to the dipolar regime, and involve cumbersome (though absolutely technically correct) theoretical formalism based on coupled Landau-Lifshitz-Gilbert and magnetostatic scalar potential equations. 

Dipole--exchange spin waves in spherical nanostructures were also studied within a microscopic lattice-spin formalism by Nguyen and Cottam, who obtained spectra and spatial mode profiles for solid, hollow, and truncated ferromagnetic spheres and there ansambles by numerical diagonalization of the corresponding quadratic bosonic Hamiltonian~\cite{NguyenCottam2008}. While this approach captures geometry-dependent spectral evolution and mode mixing, it does not isolate the specific mechanisms responsible for the coupling between modes or separate the contributions of the different interaction regimes in a systematic way.

Exchange modes in finite nanocylinders were calculated by Lim and Garg \cite{lim2021ferromagnetic}, but without a systematic explanation of spectral degeneracies. Dipole–exchange spin-wave modes in infinite cylindrical nanowires were numerically and semi-analytically investigated by Rychły, Krawczyk, et al.\cite{rychly2019spin}. However, in contrast to spherical systems, the dipole–exchange spin-wave mode structure in finite nanocylinders remains significantly less understood. The only systematic theoretical study of it  was conducted  by Puszkarski, Krawczyk, and Lévy \cite{puszkarski2007purely}.  In their work, the authors investigated the effects of introducing exchange interaction into purely dipolar modes and the dependence of mode structure and corresponding spectra on effective exchange stiffness. Their semi-analytical approach involved decomposing the cylinder into infinitesimally thin disks, each coupled to its nearest neighbors via exchange interaction and to all others via dipolar interaction. They considered only radially-homogeneous modes.

In this work, we present a systematic analysis of 
spin-wave modes in  spherical and cylindrical magnetic  resonators composed of yttrium iron garnet (YIG) as a typical optomagnonic material are assumed to have a uniform equilibrium magnetization state. We focus on the crossover from the exchange dominant regime to the dipole dominant regime and trace the evolution of spin-wave modes, noting the lifting of mode degeneracy with respect to orbital angular momentum values and the consequent mode hybridization effects, as  illustrated schematically  in Fig.~\ref{fig:Main}.      The characteristic sizes of the  structures under study vary in the range from nanometers to micrometers. In Sec.~\ref{sec:framework}, we present the complete theoretical framework based on the Landau–Lifshitz–Gilbert (LLG) equation. Sections ~\ref{sec:sphere} and ~\ref{sec:cyl} examine spherical and cylindrical geometries, respectively, across the exchange-dominated, dipole–exchange, and dipole-dominated regimes. For each regime, we analyze the underlying symmetries and derive the corresponding conserved quantities, emphasizing which of them remain valid during dynamical crossovers between the regimes. 
Then, we explain the origin of the degeneracy of the exchange spin modes, as discussed in Secs.\ref{sec:sphere-exch-model} and \ref{sec:cyl-exchange}.
Based on the obtained results, in Sec.~\ref{sec:cmt}, we introduce a coupled-mode approach to  describe distinct dipole-exchange spectral features, such as the hybridization of spin-wave modes and the anticrossing of energy terms.  Finally, in Sec.~\ref{sec:Discussion} we discuss the non-uniform equilibrium magnetization states that arise in both resonator geometries and determine the critical external field and system sizes required to obtain a uniform stationary magnetization.

\section{Theoretical framework}
\label{sec:framework}
\subsection{Governing equations}

We consider magnetization dynamics in a ferrimagnetic body occupying a volume $V$ with boundary $\partial V$.
The starting point is the LLG equation for the magnetization field
$\mathbf{M}=\mathbf{M}(t,\mathbf{r})$\cite{Akhiezer1967},
\begin{equation}\label{eq:LLG-general-form}
    \partial_t \mathbf{M}
    = -\gamma\,\mathbf{M}\times\mathbf{H}^{\mathrm{eff}}
    + \frac{\lambda}{M_s}\,\mathbf{M}\times \partial_t\mathbf{M},
\end{equation}
where $\gamma$ denotes the gyromagnetic ratio divided by the vacuum magnetic permeability, $\lambda$ is the Gilbert damping constant, and $M_s$ is the saturation magnetization. We are interested in spin-wave spectra of low-loss resonant systems, therefore we omit Gilbert damping term at this stage. The effective field $\mathbf{H}^{\mathrm{eff}}$ collects the relevant micromagnetic contributions,
\begin{equation}\label{eq:Heff-decomposition}
    \mathbf{H}^{\mathrm{eff}}
    = \mathbf{H}^{\mathrm{ext}}
    + \mathbf{H}^{\mathrm{ani}}
    + \mathbf{H}^{\mathrm{exch}}
    + \mathbf{H}^{\mathrm{dm}}
    \, ,
\end{equation}
where $\mathbf{H}^{\mathrm{ext}}$ is the applied static field,
$\mathbf{H}^{\mathrm{exch}}$ is the exchange field, and $\mathbf{H}^{\mathrm{dm}}$ is the demagnetizing field. In what follows, we neglect the crystalline anisotropy (and other symmetry-breaking interactions such as Dzyaloshinskii–Moriya terms) and treat the exchange as isotropic, since our goal is to isolate the fundamental spin-wave dynamics. These additional contributions, however, can be incorporated straightforwardly within the coupled-mode framework developed later as extra effective-field (potential) terms (see Sec.\ref{sec:Discussion}). 

An effective \textit{exchange field} $\mathbf{H}^{\mathrm{exch}}$ describes short-length spin correlations in classical continuum  models. For an isotropic exchange stiffness constant $A$, which is a reasonable approximation for cubic lattices \cite{Klingler_2015}, the exchange field can be written as 
\begin{equation}\label{eq:exchange-field}
   \mathbf{H}^{\mathrm{exch}} = \frac{A}{M_s}\nabla^2 \ \mathbf{M},
\end{equation}
where $\nabla^2$ denotes the vector Laplacian acting component-wise.
Overall prefactors depend on the unit convention. { We define the exchange constant as $A = 2D/(\mu_0 M_s)$, where $D$ is the exchange stiffness. In SI units, $A$ therefore has dimensions of $[\si{A\cdot m}]$.}

The demagnetizing field captures the long-range \textit{dipolar interaction} which also can be understood as non-local effective response to the magnetization $\mathbf{M}$.
In the magnetostatic approximation \cite{rezende2020fundamentals}, it is expressed via a scalar potential $\Psi$,
\begin{equation}\label{eq:dm-from-potential}
    \mathbf{H}^{\mathrm{dm}}(t,\mathbf{r}) = -\nabla \Psi(t,\mathbf{r}),
\end{equation}
which satisfies the Poisson equation inside the magnetic body,
\begin{equation}\label{eq:poisson-Psi}
    \nabla^2 \Psi(t,\mathbf{r}) = \nabla\cdot\mathbf{M}(t,\mathbf{r})
    \qquad (\mathbf{r}\in V),
\end{equation}
together with the standard boundary conditions at $\partial V$ Eq.~(\ref{eq:potential-d-continuity}).
An equivalent and  more convenient for our purposes representation is the integral form
\begin{equation}
\label{eq:dm-integral-kernel}
    \mathbf{H}^{\mathrm{dm}}(t,\mathbf{r})
    = \int_V{d^3r'\,
    \hat{K}(\mathbf{r}-\mathbf{r}')\cdot \mathbf{M}(t,\mathbf{r}')},
\end{equation}
with the convolution dipolar kernel  $\hat{K}(\mathbf{r}-\mathbf{r}')$ being a Hessian of the Green's function of Eq.~\eqref{eq:poisson-Psi}
\begin{equation}\label{eq:dipole-convolution-kernel}
\begin{split}
    \hat{K}(\mathbf{r}-\mathbf{r}')
    = -\nabla \nabla  \frac{1}{4\pi|\mathbf{r}-\mathbf{r}'|} =  \\ \frac{\mathbb{I}}{|\mathbf{r}-\mathbf{r}'|^3}\ - 3\frac{(\mathbf{r}-\mathbf{r}')\otimes (\mathbf{r}-\mathbf{r}')}{|\mathbf{r}-\mathbf{r}'|^5}
\end{split}
\end{equation}
where $\mathbb{I}$ is the  unity tensor and  "$\otimes$" is the tensor product.

\subsection{Linearization around the equilibrium state}

We use standard linearizarion procedure, expanding the dynamic magnetization vector around the static equilibrium magnetization $\mathbf{M}_0(\mathbf{r})$ by substituting
\begin{equation}\label{eq:linearization-substitution}
\begin{split}
     &\mathbf{M}(t,\mathbf{r}) = \mathbf{M}_0(\mathbf{r}) + \mathbf{m}(t,\mathbf{r}) , \\
     &\mathbf{H}^{\mathrm{eff}}(t,\mathbf{r}) = \mathbf{H}^{\mathrm{eff}}_0(\mathbf{r}) +  \mathbf{h}^{\mathrm{eff}}(t,\mathbf{r}) .
\end{split}
\end{equation}
where $\mathbf{m} $ and $\mathbf{h}$ are small magnetization and magnetic field deviation vectors respectively. This deviation vectors are describing precession of the total magnetization, and are subject to linearization constraints 
\begin{equation}
    \label{eq:linearization-constraints}
    {|\mathbf{m}| \ll |\mathbf{M}_0| = M_s, \qquad 
    |\mathbf{h}^{\mathrm{eff}}| \ll |\mathbf{H}^{\mathrm{eff}}_0|,}
\end{equation}
so in the linearized theory the dynamical magnetization $\mathbf{m}$ is transverse to the equilibrium magnetization,
$\mathbf{M}_0\cdot\mathbf{m}=0$, as follows from the conservation of the magnetization modulus. The perturbation of the effective field $\mathbf{h}^{\mathrm{eff}}$ is assumed to be small and linear in $\mathbf{m}$, and can be decomposed into the exchange and dipolar
contributions
$\mathbf{h}^{\mathrm{eff}}=\mathbf{h}^{\mathrm{exch}}+\mathbf{h}^{\mathrm{dm}}$ and, in particular, $\mathbf{h}^{\mathrm{exch}}=(A/M_s)\cdot\nabla^2\mathbf{m}$, while $\mathbf{h}^{\mathrm{dm}}=-\nabla\psi$ is obtained from Eqs.~(\ref{eq:dm-from-potential}--\ref{eq:poisson-Psi}) and with the scalar potential perturbation $\psi(\mathbf{r})$ is induced by the magnetization deviation $\mathbf{m}$. Substituting Eq.~\eqref{eq:linearization-substitution}  into the LLG and the Poisson equations, we retrieve the governing equations for the  equilibrium configuration
\begin{equation}\label{eq:equilibrium-condition}
    \mathbf{M}_0\times \mathbf{H}^{\mathrm{eff}}_0 = 0.
\end{equation}
In the general case, equilibrium magnetization can be an arbitrary non-uniform vector field which distribution within the micromagnetic structure volume depends on the geometry, external fields, and material parameters. However, for many micromagnetic geometries a uniform equilibrium magnetization approximation can be valid. For the case of a spherical geometry it is valid rigorously \cite{Akhiezer1967}. We discuss critical system sizes and external fields allowing uniform magnetization for the considered spherical and cylindrical geometries in Section \ref{sec:Discussion}.

Keeping the linear terms with respect to $\mathbf{m}$ in the LLG equation, yields coupled system for $\mathbf{m}$ and the dynamic magnetostatic potential $\psi$,
\begin{align}
\label{eq:linear-LLG-system}
    -\frac{1}{\gamma}\,\partial_t\mathbf{m}
    &= \mathbf{M}_0\times \mathbf{h}^{\mathrm{eff}}
      + \mathbf{m}\times \mathbf{H}^{\mathrm{eff}}_0,
      \\
\label{eq:linear-scalar-potential-system}
    \nabla^2\psi(t,\mathbf{r})
    &= \nabla\cdot\mathbf{m}(t,\mathbf{r})
    \qquad (\mathbf{r}\in V).
\end{align}

In the further, we will be interested in the eigen modes analysis, thus the frequency domain is assumed
$\mathbf{m}(t,\mathbf{r})=\mathbf{m}_\nu(\mathbf{r})e^{-i\Omega_\nu t}$ and $\psi(t,\mathbf{r})=\psi_\nu(\mathbf{r})e^{-i\Omega_\nu t}$.
Equations~\eqref{eq:linear-LLG-system}--\eqref{eq:linear-scalar-potential-system} then define an eigenvalue problem for $\Omega_\nu$ and the mode profiles $\{\mathbf{m}_\nu,\psi_\nu\}$ is subject to exchange and magnetostatic boundary conditions at $\partial V$
\begin{subequations}\label{eq: BC}
    \begin{equation}
    \label{eq:potential-continuity}
        \psi\big\rvert_{\partial V+0}=\psi\big\rvert_{\partial V - 0} 
    \end{equation}
    \begin{equation}\label{eq:potential-d-continuity}
        \mathbf{n}_{\partial V}\cdot\left(\frac{\partial\psi}{\partial \mathbf{r}}\bigg\rvert_{\partial V + 0}-\frac{\partial\psi}{\partial \mathbf{r}}\bigg\rvert_{\partial V - 0}\right) = \mathbf{n}_{\partial V}\cdot\mathbf{m}
    \end{equation}
    \begin{equation} \label{eq:magnetization-neumann}
        \mathbf{n}_{\partial V}\cdot\frac{\partial\mathbf{m}}{\partial\mathbf{r}}\bigg\rvert_{\partial V} = 0
    \end{equation}
\end{subequations}

Equations (\ref{eq:linear-LLG-system} -\ref{eq:linear-scalar-potential-system}) equipped with boundary conditions Eq.~(\ref{eq:potential-continuity}-\ref{eq:magnetization-neumann}) provide comprehensive description of any spin-wave resonant system. 

{However, this approach that couples the LLG with the scalar magnetic potential offers full compact system of equation, it can be further reduced to integro-differential equation  formulated in  magnetization-only terms and which will help further analysis. To eliminate the scalar potential, we employ integral form of the demagnetizing field Eq.~\eqref{eq:dm-integral-kernel} with kernel Eq.~\eqref{eq:dipole-convolution-kernel}.
We substitute $\mathbf{h}^{\mathrm{eff}}$ into equation Eq.~\eqref{eq:linear-LLG-system} and rewrite the dynamic demagnetizing field in the integral form to obtain the integro-differential governing equation

\begin{widetext}
\begin{equation}\label{eq:linear-LLG-full}
    -\partial_t\mathbf{m}(t,\mathbf{r}) = \underbrace{\gamma\frac{A}{M_s}\mathbf{M}_0\times\nabla^2\mathbf{m}(t,\mathbf{r})}_{\text{I: exchange interaction term}} - \underbrace{\gamma\left( \mathbf{H}^{\mathrm{ext}} + \mathbf{H}^{\mathrm{dm}}_0 \right)\times\mathbf{m}(t,\mathbf{r})}_{\text{II: Larmor precession term}} + \underbrace{\gamma\mathbf{M}_0\times\int_{V}{d^3\mathbf{r}' \ \hat{K}(\mathbf{r}-\mathbf{r}')\cdot\mathbf{m}(t,\mathbf{r}')}}_{\text{III: dipole interaction term}}.
\end{equation}
\end{widetext}
Note that due to  uniform magnetization aligned with $z$-axis static effective magnetic field does not contain exchange term as $\nabla^2\mathbf{M}_0 = 0$ and deviations vector has only two (orthogonal to equilibrium) components $\mathbf{m} = (m_x, m_y, 0)$. 
Static demagnetizing field $\mathbf{H}_{0}^{\mathrm{dm}}$, however, is given by the integral Eq.~\eqref{eq:dm-integral-kernel} with $\mathbf{M}=\mathbf{M}_0$, and while it is uniform in sphere $\mathbf{H}_0^{\rm{dm}}=-1/3 \ \mathbf{M}_0$, it is generally non-uniform in a finite cylinder. 

\subsection{Interaction regimes}
Equation \eqref{eq:linear-LLG-full} is convenient not only because it involves only spin-wave modes as the unknown, but also because it makes different interactions mechanisms explicit by assigning them different linear operators. From this formulation, we distinguish three limiting regimes that will be used throughout the paper.

\paragraph*{Exchange-dominated regime.}
For sufficiently short-wavelength excitations or for samples of small spatial extent, the dynamics is governed mainly by \textit{exchange interaction} term, while the \textit{dipole interaction} term can often be neglected. In this limit, Eq.~\eqref{eq:linear-LLG-system} reduce to a local partial differential equation  and provide an eigenmode basis corresponding to the exchange spin-wave solutions.

\paragraph*{Dipole-dominated (magnetostatic) regime.}
For long-wavelength excitations in sufficiently large samples, the exchange term becomes subleading and the dynamics is governed primarily by the non-local dipolar interaction.
In this limit, one recovers the classical magnetostatic (Walker-type) modes where the \textit{dipole interaction} governs the system’s behavior, turning Eq.~\eqref{eq:linear-LLG-full}
into Fredholm kind II integral equation, with the mode structure is set by sample shape and external field.

\begin{figure*}[t]
    \centering
\includegraphics[width=1\textwidth]{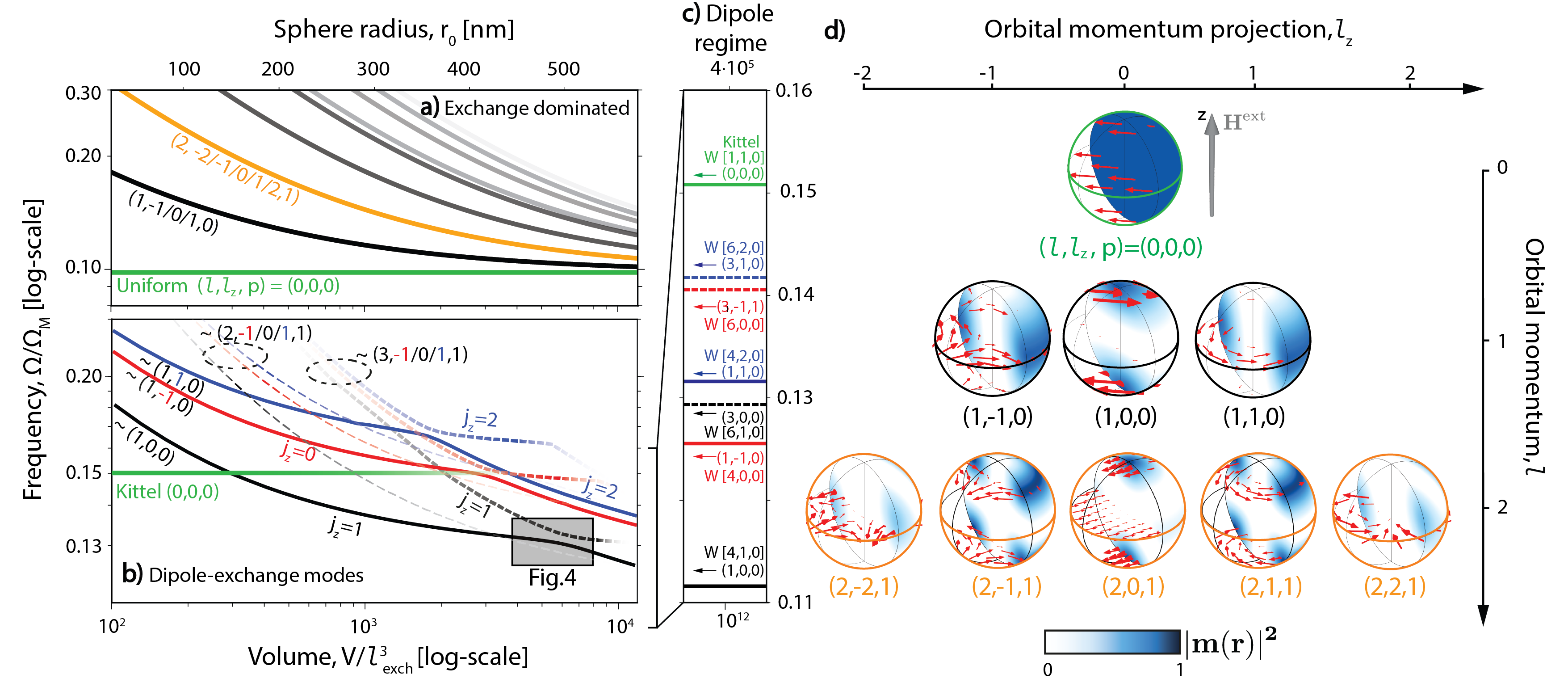}
    \caption{ \label{fig:Sphere}
    Spin-wave spectrum of a uniformly magnetized sphere as a function of volume $V$ normalized by the exchange volume $l_{\mathrm{exch}}^3$, shown in three interaction regimes: (a) exchange, (b) dipole-exchange, and (c) dipole. (a) Exchange-dominated regime consists of degenerate multiplets characterized by $(l,l_z,p)$ denoting the orbital angular momentum, its projection, and the radial number. The uniform mode $(0,0,0)$ is highlighted in green. (b) Dipole--exchange regime, the nonlocal dipolar interaction lifts the exchange degeneracy and induces hybridization between modes \roundtwo{from the same symmetry sector characterized by $j_z$ denoting the total angular momentum projection} (gray rectangle; see FIG.\ref{fig:CMT}).
The Kittel mode does not hybridize with other modes, as discussed in Sec.\ref{subsec:cmt:c}. Only a subset of modes is shown to improve readability.
(c) Dipole dominated regime shows the continuation of the dipole–exchange modes for large volume, where the spectrum evolves into the well-known Walker-mode structure characterized by the indices $[n,m,r]$\cite{walker1958,fletcher1959}. (d) Representative mode profiles. Spatial distributions of $|\mathbf m(\mathbf r)|^2$ and dynamical magnetization flow (red arrows) are shown for lowest-energy exchange modes from panel a). In all panels  the external field is ${\mathbf H^{\mathrm{ext}} = 13.3 \cdot 10^4 \si{A/m} = 1.6\cdot 10^3 \ \si{Oe}}$ and is applied along $\mathbf{z}$ axis.}  
\end{figure*}
\paragraph*{Dipole-exchange regime.}
The most complex phenomena appear in the intermediate regime, where neither interaction can be neglected. The resulting eigenmodes are hybrid dipole-exchange spin-wave modes, whose spectra interpolate continuously between the two limiting cases. This is the regime where the dynamics is richest, allowing for mode hybridization and nontrivial coupling phenomena.

The relative importance of exchange and dipolar interactions depends on the characteristic length scales (e.g., cavity size $L$ and/or the characteristic wave number $k$ of a mode).
A useful material length scale is the exchange length
\begin{equation}\label{eq:exchange-length}
    \ell_{\mathrm{exch}} = \sqrt{\frac{A}{M_s}}
    \, ,
\end{equation}
where $A$ is the (scalar) exchange {constant} in the isotropic limit.

\subsection{Details of computations}
\label{sec:computation-details}

The material assumed throughout the paper is YIG with exchange stiffness $D=1.29 \cdot 10^{-11}~\si{J/m} = 1.29\cdot 10^{-6} \ \si{erg/cm}$ \cite{deb2019damping}, saturation magnetization $ M_s = 1.40 ~\cdot~10^5~\si{A/m} = 1.40~\cdot~10^{2} \ \si{emu/cm^3}
$ \cite[p.~333]{daniel2009spin}, gyromagnetic ratio {$\gamma=2.21\cdot 10^5 \ \si{m/(A\cdot s)} = 22.1 \   \si{G^{-1} s^{-1}}$}. This give the value of the exchange length $\ell_{\mathrm{exch}}\approx 32$ nm. The external field was also set as constant; for the sphere, it was set to $H^{\rm{ext}}=13.3 \cdot 10^4 \si{A/m} =1.6\cdot 10^3 \ \si{Oe} $, and for cylinder calculations, it is {$H^{\rm{ext}}=5.5 \cdot 10^4 \si{A/m} = 0.7\cdot 10^3 \si{Oe}$.}

{Equation \ref{eq:linear-LLG-full} was solved numerically} using the finite-element method to obtain spin-wave spectra of a sphere and a cylinder presented in figures \ref{fig:Sphere} and \ref{fig:Cylinder} respectively. 
Dipole-exchange modes were computed with the finite-element package COMSOL Multiphysics, solving linearized LLG Eq.~\eqref{eq:linear-LLG-system} with the help of Micromagnetics module \cite{10.1063/5.0143262,comsol}  coupled to the Poisson equation for scalar potential Eq.~\eqref{eq:linear-scalar-potential-system} with Magnetic Field, No Currents (MFNC) module. The static demagnetizing field $\bf M_0 (\bf r)$ was also computed with MFNC. 
Modes in the dipole-interaction limit were calculated using internally developed discrete dipole solvers, which approximates integral convolution Eq.~\eqref{eq:dipole-convolution-kernel} with a square matrix on the COMSOL generated mesh. The code can be found in the GitHub repository \url{https://github.com/aluatar/integsol.git} 
 
\section{Spin-wave Modes of a Sphere}

\label{sec:sphere}
Let us first consider a 
{primer}
case of spin-wave modes in a sphere. 
While the overview of the consecutive studies of this problem has been  discussed in Section~\ref{sec:intro}, we provide here a comprehensive and solid digest of the theoretical approach based on LLG solution. 

\subsection{Exchange interaction. Angular momentum picture.}
\label{sec:sphere-exch-model}

We start with the exchange-dominated regime when the characteristic spatial variation scale of the dynamical magnetization is sufficiently short so that the non-local dipolar interaction can be neglected to leading order. Physically, this corresponds to eigenmodes with a characteristic wave number~$k$ satisfying 
\begin{equation}
k\,\ell_{\mathrm{exch}} \gg 1,
\label{eq:exchdom_condition}
\end{equation}
where $\ell_{\rm{exch}}$ denotes the exchange length as specified in Eq.~\eqref{eq:exchange-length}. In this regime, the restoring dynamics of spin-wave excitations are primarily governed by the exchange interaction, and the dipolar interaction enters via the static demagnetizing field that defines the equilibrium magnetization configuration. Therefore,  Eq.~\eqref{eq:linear-LLG-full} for dynamic magnetization reduce to 
\begin{equation}
    \label{eq:linear-LLG-exch}
   -\partial_t\mathbf{m}(t,\mathbf{r}) =\gamma \big(\ell_{\mathrm{exch}}^2\mathbf{M}_0\times\nabla^2- \left( \mathbf{H}^{\mathrm{ext}} + \mathbf{H}^{\mathrm{dm}}_0 \right)\times\big)\mathbf{m}(t,\mathbf{r}).
\end{equation}
Writing the right-hand side of  Eq.~\eqref{eq:linear-LLG-exch} as a linear operator $\mathcal{H}_{\text{exch}}$, one can introduce {Hermitian} effective Hamiltonian acting on the spin-wave field
\begin{gather}
\label{eq:Schroedinger}
    -\textit{i}\partial_t\mathbf{m}(t,\mathbf{r}) = \mathcal{H}_{\text{exch}} \ \mathbf{m}(t,\mathbf{r}).
\end{gather}
Therefore, by applying this equation, we seek the eigenmodes of the spherical spin-wave resonator in the exchange-dominated regime.

\emph{Angular momentum picture and symmetry considerations.} Before turning to providing rigorous solution of this problem, we need to highlight that the problem has particular symmetries which immediately provide the key integrals of the problem. First of all, despite that the resonator has spherical symmetry, the static magnetic field induces equilibrium homogeneous magnetization  $\mathbf{M}_0,\mathbf{H}_0^{\mathrm{eff}} \parallel \mathbf{\widehat{z}}$, has reducing the symmetry to cylindrical.  Hence,  $\mathbf{m} = (m_x,m_y)\in \mathbb{R}^2$  has only two components with $SO(2)$ symmetry. Considering the action of the $SO(3)$ group on the spin-wave vector field
\begin{gather}
\label{eq:rotation}
    \mathbf{m}(\mathbf{r})\overset{R\in SO(3)}{\rightarrow} R\ \mathbf{m} (R^{-1}\mathbf{r})=\mathbf{m'}(\mathbf{r'}), 
\end{gather}
natural definition of angular momenta can be introduced. Indeed,  $\mathbf{m}(\mathbf{r})$ is  embedded in $\mathbb{R}^3$ and rotations of the components of vector field is generated by spin-1 operators $\widehat{S}_m, m\in\{x,y,z\}$,   
\begin{equation}
    R=e^{-i\alpha (\mathbf{n}, \mathbf{\widehat{S}})} \qquad (S_k)_{jn} = i \varepsilon_{jkn},
\end{equation}

with $\varepsilon_{ijk}$ is a Levi-Civita symbol, and spatial rotations are generated by orbital angular momentum (OAM) operator $\widehat{\mathbf{{L}}}=-{i} \mathbf r \times \mathbf{\nabla}$. { Here $(\mathbf{n}, \mathbf{\widehat{S}}) = n_i(S_i)_{jk}$}. Thus, the $SO(3)$ rotation Eq.~\ref{eq:rotation} turns to 
\begin{gather}
    \mathbf{m'}(\mathbf{r'}) = e^{-i\alpha(\mathbf{n},\mathbf{\widehat{S}})}e^{-i\alpha(\mathbf{n},\mathbf{\widehat{L}})}\mathbf{m}(\mathbf{r})= e^{-i\alpha(\mathbf{n},\mathbf{\widehat{J}})}\mathbf{m}(\mathbf{r}),
\end{gather}
where $\mathbf{\widehat{J}}=\mathbf{\widehat{S}}+\mathbf{\widehat{L}}$ is the total angular momentum. 

The problem is symmetric with respect to a particular transformation if it (or its generator) commutes with the Hamiltonian. Therefore, to establish symmetries and related conserved quantities, we find which generators of total rotations or its components Eq.~\eqref{eq:rotation} commute with the effective exchange Hamiltonian of Eq.~\eqref{eq:Schroedinger}.

Using the identity $\mathbf a\times \mathbf b=-i(\widehat{\bm{S}}\cdot \bf{a})\bf{b}$, terms of Eq.~\eqref{eq:linear-LLG-exch} can be rewritten as
\begin{gather}
   \mathbf{M}_0\times\nabla^2\mathbf{m} = -\textit{i}(\mathbf{M}_{0}\cdot\widehat{\mathbf{S}})\ \nabla^2\mathbf{m} = -\textit{i}M_s\widehat{S}_z\nabla^2\mathbf{m},\notag\\
\mathbf{H}_0^{\text{eff}}\times\mathbf{m} = -\textit{i}(\mathbf{H}_0^{\text{eff}}\cdot\widehat{\mathbf{S}})\ \mathbf{m} = -\textit{i}H_{0}^{\text{eff}}\widehat{S}_z \ \mathbf{m},
\end{gather}
where $\mathbf{H}_{0}^{\text{eff}}= \mathbf{H}^{\mathrm{ext}} + \mathbf{H}^{\mathrm{dm}}_0$. Then the Hamiltonian from Eq.~\ref{eq:Schroedinger} takes form 
\begin{gather}
\label{eq:1lo}
   \mathcal{H}_{\text{exch}} = \gamma\widehat{S}_z\cdot\left(\ell_{\text{exch}}^2 \ M_s\nabla^2 - H_0^{\text{eff}}\right)\mathbb{I}.
\end{gather}
The operator ${\nabla}^2$ is \emph{vector Laplacian}, which in Cartesian components takes diagonal form

\begin{gather}
\label{eq:laplacian}
    \nabla^2 = \Delta\mathbb{I} \notag \\
    \Delta = \Delta_{r}-\frac{\widehat{L}^2}{r^2} = \Delta_{r,\theta} -\frac{1}{r^2\sin^2\theta}\widehat{L}_z^2,
\end{gather}
with $\Delta$ being a scalar  Laplacian written in the spherical coordinates. 

One can immediately see that the Hamiltonian commutes with $\widehat{L}^2$
\begin{equation}
[\widehat{L}^2,\mathcal{H}_{\text{exch}}] =0,
\end{equation}
as it commutes with the radial part of the Laplacian and constant vectors. Also, the Hamiltonian commutes with orbital angular momentum components $\hat{L}_i$, $i\in (x,y,z)$ (See Appendix~\ref{appx:spin-1-operators})
\begin{equation}
    \label{sphere-Hamiltonian-Lz-commutation}
    [\widehat{L}_i,\mathcal{H}_{\text{exch}}] = 0,
\end{equation}
which means that the system is invariant under coordinate rotations.
Since the Hamiltonian also evidently commutes with  $\widehat{S}_z$, finally,
\begin{equation}
    [\widehat{L}_z + \widehat{S}_z, \mathcal{H}_{\text{exch}}] = [\widehat{J}_z, \mathcal{H}_{\text{exch}}] = 0  
\end{equation}

Thus, these operators set a proper basis for labeling the modes of {\it exchange-dominated} regime. However, as we will show later, after retaining the non-local dipole interaction, only $\widehat{J}_z$ will remain a conserved quantity.

Finally, we observe that the spherical problem is invariant under a set of discrete symmetries, specifically under spatial inversion of the coordinates, $\hat{\mathcal{I}}\mathbf{r} \to -\mathbf{r}$, and under the { full reflection,} in the $xOy$-plane, $\hat{P}_z$, which reverses both the $z$ coordinate axis and the $z$ components of vectors simultaneously. 

{\it Rotating basis}.  Since the dynamic magnetization vector manifests itself with precession, it is very convenient to translate the $\mathbf{m}$ from Cartesian to circular basis:
\begin{equation}
\label{eq:transform-cart-circ1}
    m_+=-\frac{1}{\sqrt{2}}(m_x + \textit{i}m_y), \ \ m_-=\frac{1}{\sqrt{2}}(m_x - \textit{i}m_y).
\end{equation}
In the following, a tilde $\tilde {\mathbf{m}}=(m_+,\, 0,\, m_-)^T$ denotes vectors and operators expressed in the circular basis. 
The transformed effective Hamiltonian Eq.~\eqref{eq:1lo}  obtains the form
\begin{align}
    \label{eq:transform-cart-circ2}
  \tilde{\mathcal{H}}_{\text{exch}}  = \mathrm{U}_{\mathrm{cart}\rightarrow \mathrm{circ}}\cdot\mathcal{H}_{\text{exch}}\cdot\mathrm{U}^{-1}_{\mathrm{cart}\rightarrow \mathrm{circ}}
\end{align}
\begin{align}
    -\textit{i}\partial_t\tilde{\mathbf{m}}(t,\mathbf{r}) = \tilde{\mathcal{H}}_{\text{exch}} \ \tilde{\mathbf{m}}(t,\mathbf{r}),\notag \\ 
  \tilde{\mathcal{H}}_{\text{exch}} = \gamma\widehat{\tilde{S}}_z\cdot\left(\ell_{\text{exch}}^2 \ M_s\tilde{\nabla}^2 - H_0^{\text{eff}}\right)\mathbb{I},\label{eq:sphere-hamiltonian-equation-in-pol-basis}
\end{align}
where $\widehat{\tilde{S}}_z$ becomes diagonal in the circular basis (See Appendix~\ref{appx:spin-1-operators}). Thus, finally, Eq.~\eqref{eq:sphere-hamiltonian-equation-in-pol-basis} can be decoupled into two independent equations 
\begin{equation}
    \textit{i}\partial_t m_{\pm} = \pm \gamma\left(\ell_{\text{exch}}^2M_s\Delta - H_0^{\text{eff}} \right)m_{\pm}
    \label{eq:sphere-exchange-governing-equation}
\end{equation}
for right-handed $m_+$ and left-handed $m_-$ circular polarizations. This equation has the structure of a scalar time-independent Schrodinger equation and, after reduction to the Helmholtz equation, permits an analytical solution.

\subsection{Rigorous solution for exchange modes}
\label{sec:rig_sphere-exch-modes}

We proceed with Eq.~\eqref{eq:sphere-exchange-governing-equation} which, after substitution of time-harmonic ansatz $m_{\pm}\propto e^{-\mathrm{i}\Omega t}$, becomes an eigenvalue problem 
\begin{equation}
    \label{eq:sphere-exchange-eigenvalue-problem}
    \Omega\ m_{\pm} = \pm\gamma (\ell_{\mathrm{exch}}^2M_s\Delta - H^{\mathrm{eff}}_0)m_{\pm},
\end{equation}
then we divide the equation by $\ell_{\mathrm{exch}}^2 \ \Omega_M$ with $\Omega_M = \gamma M_s$, rearrange the terms to bring it to a Helmholtz-type eigenproblem 
\begin{equation}
\Big(\Delta + k^2\Big)\,m_{\pm}= 0,
\qquad
k^2 = \frac{1}{\ell_{\text{exch}}^2}\frac{\mp\Omega - \Omega_K}{\Omega_M},
\label{eq:exchdom-helmholtz}
\end{equation}
where {$\Omega_{\text{uni}}= \gamma (H^{\text{ext}}- 1/3M_s)$ is the uniform mode frequency}. Here we used the explicit formula for the static demagnetizing field in the uniformly magnetized sphere $H^{\mathrm{dm}}_0 = -1/3 M_s$. 
The resulting eigenproblem reduces to the spherical Helmholtz equation for the polarization amplitudes $m_\pm(\mathbf r)$, supplemented by the regularity condition $|m_\pm|<\infty$ at $r=0$ and the boundary condition Eq.~\eqref{eq:magnetization-neumann} which takes the form
\begin{equation}
    \label{eq:sphere-neumann-boundary-condition}
    \left.\frac{d m_{\pm}}{dr} \right|_{r=r_0}=0,
\end{equation}

The general form of the solution is standard 
\begin{equation}
\label{eq:sphere-exchange-eigenmode}
m_{\pm,l,l_z}(r,\theta,\varphi)=C_{l, l_z}\, j_l(k r)\,Y_l^{l_z}(\theta,\varphi),
\end{equation}
where $j_l(x)$ is the spherical Bessel function of the first kind, and $Y_l^{l_z}(\theta,\varphi) $ is the spherical harmonic. Here, numbers $l$ and $l_z$ are associated with the orbital angular momentum and its projection on the $z$-axis. 

Eigenmodes given by Eq.~\eqref{eq:sphere-exchange-eigenmode} are admissible for both polarizations, $m_+$ and $m_-$. However, in the case of positive polarization $m_+$, saturated magnetization regime $\Omega_K>0$, and  with our definition of the wave number $k$ in Eq.~\eqref{eq:exchdom-helmholtz}, the relation $k^2(\Omega)>0$ is satisfied only for $\Omega<-\Omega_K$, whereas $\Omega>0$ would require an imaginary radial wavenumber $k$, which is incompatible with the boundary conditions. As a result, for the $m_+$ polarization, the associated eigenfrequency appears negative, backward-in-time-propagating solutions.
In the following we therefore work with the $m_-$ polarization as the positive frequency branch, while keeping in mind that the complementary $m_+$ sector remains part of the complete eigenbasis and plays the role of the time-reversed (backward-propagating) partner required by time-reversal symmetry and participates in polarization mixing in dipole-exchange and dipole regimes, as discussed in Sec.~ \ref{sec:Discussion}.

Finally, by satisfying the radial boundary conditions, one gets the full set of quantum numbers of the systems, and $m_{-,l,l_z,p}\sim j_l(k_{l,p} r)\,Y_l^{l_z}(\theta,\varphi)$, where $p$ denotes the order of the root of the derivative of the spherical Bessel function, $j_l'(\kappa)=0$, and $k_{l,p}={\kappa_{l,p}}/{r_0}$, and root numeration starts from zero. The normalization constant $C_{l, l_z, p}$ can be found from the condition $\int_V|m_-(\mathbf r)|^2dV=1$.

The resulting eigenfrequencies as a function of the sphere's radius and the external field are given by a simple relation:
\begin{equation}
    \label{eq:pure-exch-sph-eigenfreq}
    \dfrac{\Omega_{l p}}{\Omega_M} = \dfrac{\kappa_{l,p}^2\ell_{\text{exch}}^2}{r_0^2}  +   \dfrac{\gamma H^{\text{ext}}}{\Omega_M}- 1/3
\end{equation}

Analyzing the obtained results, firstly, one can see that the spin-wave spectrum is degenerate with respect to the orbital angular momentum projection $l_z$. For each fixed orbital index $l$ the eigenfrequencies form the standard $2l+1$ multiplet with $l_z = -l,...,l$ typical for a spherically symmetric problem. At first glance, such spherical degeneracy appears surprising because the equilibrium magnetization selects a preferred axis, which would naively suggest lifting of the spherical multiplet structure. However, since the equilibrium magnetization is homogeneous, this results in lifting only the spin part. Indeed, by changing from Cartesian to circular basis $m_{x,y}\to m_{\pm}$ one factorizes the spin part, since $\widehat{\tilde{S}}_z m_{\pm}=\mp m_{\pm}$. Moreover,  as it was previously discussed, these states can also be labeled by total angular momentum projection $\widehat{\tilde{J}}_z=\widehat{\tilde{L}}_z + \widehat{\tilde{S}}_z$ and $\widehat{\tilde{J}}_z m_-=j_z m_-=(l_z+1)m_-$ so  both representations are equivalent.  This degeneracy is clearly seen in Fig.~\ref{fig:Sphere} (a), where the states corresponding to the exchange dominated interaction are showed. We label the states by $(l,l_z,p)$ quantum numbers. 

Secondly,  one can notice that eigenfrequencies are inversely proportional to the the square of the the sphere's radius, which results in a rapid increase in the eigenfrequencies  for small enough radii, as captured in Figure~\ref{fig:Sphere}(a). At the same time, there is Kittel mode (0,0,0) with dynamical magnetization uniformly precessing across the sphere volume at frequency $\Omega_{000} = \Omega_K = \gamma(H_0-1/3M_s)$. In the limiting case of a very large sphere $r_0/\ell_{\text{exch}}\to\infty$, all the frequencies tend to the Kittel frequency, thus illustrating that exchange interaction lifts the degeneracy only for radii comparable to the exchange length.

Finally, several lowest-frequency mode profiles are shown in Fig.~\ref{fig:Sphere}(d) for $l=0\ldots 2$ and various $l_z$ and $p$ numbers, showing the structure of the dynamical magnetization vector. The higher-order modes are indicated by faint gray lines in Fig.~\ref{fig:Sphere}(a).

\subsection{Dipole--exchange modes}
\label{sec:sphere-dipole-exchange}

The pure-exchange description of spin-wave modes in a sphere applies only in the short-wavelength limit
$k\ell_{\mathrm{exch}}\gg1$.
Except for the spatially uniform Kittel mode ($k=0$), this condition is rapidly violated for low-order modes as the sphere radius increases.
For experimentally relevant radii $r_0\gtrsim 100$ nm, the system therefore enters the dipole–exchange regime, where the long-range dipolar interaction must be treated on equal footing with the exchange.

In this regime, the linearized dynamics is governed by the effective Hamiltonian
\begin{equation}
\mathcal{H}_{\text{dip-exch}} =  \mathcal{H}_{\mathrm{exch}} + \mathcal{H}_{\rm{dip}} ,
\end{equation}
where $\mathcal{H}_{\text{exch}}$ is the local exchange operator derived in Sec.~\ref{sec:sphere-exch-model}, and the action of dipolar non-local operator on spin-wave field is given by integral convolution
\begin{equation}
(\mathcal{H}_{\rm{dip}}\,\mathbf m)_i(\mathbf r)
=
\gamma M_s \widehat{S}_{z_{,}ij}
\int_V d^3\mathbf r'\;
\widehat K_{jk}(\mathbf r-\mathbf r')\, m_k(\mathbf r').
\label{eq:K-maintext}
\end{equation}
Here $\widehat K_{jk}$ is the magnetostatic kernel and $\widehat S_z$ originates from the term
$\mathbf M_0\times\mathbf h^{\mathrm{dm}}$, we continue to consider equilibrium homogeneous magnetization  $\mathbf{M}_0 \parallel \mathbf{\widehat{z}}$.
The appearance of $\mathcal{H}_{\rm{dip}}$ qualitatively changes the symmetry structure of the eigenproblem and explains the evolution of the spectrum shown in Fig.~\ref{fig:Sphere} (b) for dipole-exchange interaction.

\paragraph{Lifting of exchange degeneracy.} In the exchange-dominated regime, spherical symmetry implies conservation of the orbital angular momentum $\widehat L^2$, leading to $(2l+1)$-fold degenerate multiplets labeled by $l_z=-l,\dots,l$.
The dipolar interaction breaks this symmetry.
In particular, the dipolar operator does not commute with the orbital angular-momentum operator,
\begin{equation}
[\widehat L^2,\mathcal{{H}_{\rm{dip}}}]\neq0 ,
\end{equation}
and therefore couples exchange modes with different orbital indices $l$, which in this regime is no longer a conserved quantity. 
As a result, the spherical symmetry protection of the exchange multiplets is lost, and the degeneracy characteristic of Fig.~\ref{fig:Sphere}(a) is progressively lifted as the system moves into the dipole--exchange regime.
This splitting of formerly degenerate branches is clearly visible in Fig.~\ref{fig:Sphere}(b).

\paragraph{Mode hybridization: anticrossings and symmetry-protected crossings.} At the same time, the dipolar interaction preserves axial symmetry with respect to the equilibrium magnetization axis. Specifically, the generator of rotations about $\widehat{\mathbf z}$ commutes with the dipolar operator,
\begin{equation}
[\widehat J_z,\mathcal{H}_{\rm{dip}}]=0 ,
\end{equation}
while neither $\widehat L_z$ nor $\widehat S_z$ separately do. 
Consequently, the eigenvalue $j_z$ of the total angular momentum operator, projected onto the $z$ axis, remains a conserved quantity in the dipole–exchange regime. {Additionally, dipolar term remains invariant under full reflection in the $xOy$ --plane $\hat{P_z}$, thus the $z$--parity is also conserved.}

Thus, Fig.~\ref{fig:Sphere}(b) reveals anticrossings in the dipole–exchange spectrum, which clearly signal mode hybridization. {Although the complete mode-coupling theory will be presented in Sec.~\ref{sec:rig_sphere-exch-modes}, the Wigner–von Neumann theorem\cite{berry1984quantal} implies that modes with different values of $j_z$ can cross without interacting, since they correspond to different representations of the symmetry group $U(1)$
generated by rotations around the $z$-axis.}
{Similarly, because the dipole--exchange operator is invariant under reflection in the $xOy$ plane (see Sec.~\ref{appx:inversion-symmetries}), modes of opposite parity do not hybridize. In spherical geometry the parity is related to the orbital and azimuthal indices as $P_z = (-1)^{l - j_z - 1}$, so that, within a given $j_z$ sector, the parity of the orbital number $l$ determines which spin-wave modes can hybridize and which remain symmetry-protected.}
While the only proper labeling of the modes should be based on $j_z$, we also label the modes in Fig~\ref{fig:Sphere} with  $(l,l_z,p)$ to show their correspondence to exchange-dominated regime {and their parity}. 

Finally, the Kittel mode having zero orbital momentum still does not couple with any other mode since for $l=0$ the commutation relation with the dipole operator still holds. This is discussed in more details in Sec.~\ref{subsec:cmt:c} and connected to Eq.~\eqref{eq:kittel-mode-dipolar-action}.

All symmetry arguments and commutator relations invoked here are derived explicitly in Appendixes \ref{appx:rotatonal-covariance-proof} and \ref{appx:dipolar-commutators}. The results of the calculations shown in Fig.~\ref{fig:Sphere} (b) were obtained with help of the numerical modeling in Comsol Multiphysics (see Sec.~\ref{sec:computation-details}). 

\subsection{Dipole interaction limit}
In the limit of large sphere size, the dipolar term dominates over the exchange part and the modes are defined by the geometry-dependent demagnetizing response rather than through the exchange-induced dispersion.  Formally, exchange can be neglected when $r_0 \gg \ell_{\mathrm{exch}}/\kappa_{l,p}$, i.e. when the particle size is large and the characteristic spatial oscillation scale set by $1/\kappa_{l,p}$ is small.

In the dipole-dominated (magnetostatic) regime, the exchange term in the master equation Eq.~\eqref{eq:linear-LLG-full} can be neglected, and the dynamics is governed by the torque in the static field  and the dipole demagnetizing interaction term: 

\begin{align}
i\Omega\mathbf m(\mathbf r)=\gamma\mathbf M_0\times \int_V{d^3\mathbf{r}',\hat{K}(\mathbf{r}-\mathbf{r}')\mathbf m(\mathbf{r}')} +\nonumber \\
+\gamma \mathbf m(\mathbf{r})\times \mathbf{H}^{\rm{eff}}_0.\label{eq:dipolar-fredholm}
\end{align}

Frequencies obtained from \eqref{eq:dipolar-fredholm} are shown in Fig.~\ref{fig:Sphere} (c). We traced these modes from the dipole–exchange regime all the way to the macroscale particle limit $r_0\sim 400\ \mu m$, so the color and line type match the corresponding  modes from Fig.~\ref{fig:Sphere} (b). One can see that in this regime the relative spectral positions of the modes change significantly, and the overall spectral range shrinks to the interval between 0.11 and 0.15.
It is also of particular interest to establish the correspondence between our results and the classical Walker–Fletcher magnetostatic theory \cite{walker1958,fletcher1959}. 
In that framework, the magnetostatic potential $\psi(\textbf{r})$ is used to construct exact eigenmodes of the form $\psi _{n,m,r}\propto r^nY_n^m(\theta ,\phi )$. The Walker notation $[n,m,r]$ labels the degree (orbital order) $n$ of the solid-harmonic expansion, the azimuthal quantum number $m$, and $r$ enumerates the roots of the characteristic equation for fixed $(n,m)$.
To relate these modes to the exchange and dipole–exchange eigenbases discussed in Secs.~\ref{sec:sphere-exch-model} and \ref{sec:sphere-dipole-exchange}, one must first obtain the dynamic magnetization via $\mathbf{m}_{n,m,r}=-\widehat{\chi} (\Omega_{n,m,r})\nabla \psi_{n,m,r}$, where $\widehat{\chi}$ is the magnetic susceptibility tensor. Although establishing a general mapping between the Walker quantum numbers and the angular-momentum labels used in our approach is not straightforward, axial symmetry ensures that the azimuthal number coincides with the total angular-momentum projection $m = j_z$.
At the same time, the number $n$ does not correspond to the orbital angular momentum $l$ since direct computation shows that each field $\mathbf{m}_{n,m,r}$ contains  three exchange harmonics with $l=n-1, n, n+1$. Still, the parity of the selected solid-harmonic degree $n$ is even or odd under $z\to -z$ within a given $m$ (or $j_z$) sector, which matches our parity label $P_z$. We have analyzed the harmonics plotted in Fig.~\ref{fig:Sphere}~(c) with different angular momenta and parity and attributed them with the Walker notation in brackets $[n, m, r]$.   

\section{Spin-wave modes of a cylinder}
\label{sec:cyl}
\begin{figure*}[t]
    \centering
    \includegraphics[width=1\textwidth]{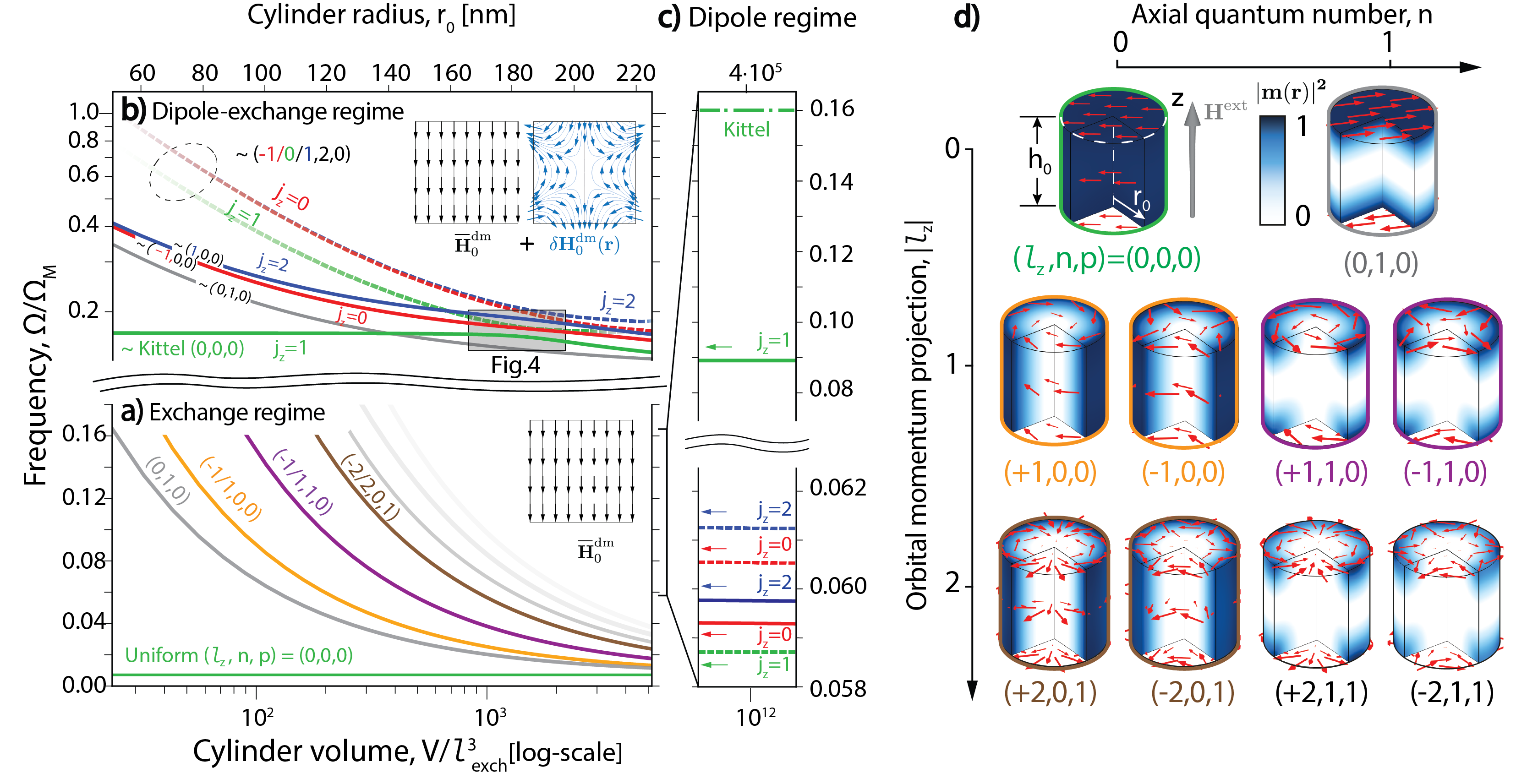}
    \caption{\label{fig:Cylinder}
    Spin-wave spectrum of ferrimagnetic cylinder as a function of volume $V$, normalized by the exchange volume $l_{\mathrm{exch}}^3$, shown in three interaction regimes: (a) exchange, (b) dipole-exchange, and (c) dipole. The top axis indicates the corresponding cylinder radius $r_0$ with fixed aspect ratio $2r_0/h_0=1$. 
(a) Exchange regime consists of degenerate multiplets labeled by $(l_z,n,p)$, denoting the orbital-momentum projection, the number of $\mathbf{z}$-axis nodes, and the radial index. Only the homogeneous demagnetizing field $\overline{\mathbf{H}}_0^{\mathrm{dm}}$(black arrows) is included.
(b) Dipole--exchange regime, the spatially varying demagnetizing field $\delta \mathbf{H_{\mathnormal{0}}^{\mathrm{dm}}}(\mathbf{r})$ (blue arrows) and the resulting nonlocal dipolar interaction $\mathcal{H}_{\rm{dip}}$ lift the exchange degeneracy and induce hybridization between modes (gray rectangle; see Fig.\ref{fig:CMT}). Only a subset of modes is shown to improve readability.
\roundtwo{The branch originating from $(l_z,n,p)=(0,1,0)$ (gray solid line) crosses the Kittel-like branch originating from $(0,0,0)$ (green solid line) without hybridization: both have $j_z=1$, but their mirror parities $P_z$ are opposite. In contrast, the $(0,0,0)$ and $(0,2,0)$ modes share both symmetry labels and hybridize in the gray rectangle [see Fig.~\ref{fig:CMT}(b)].}
(c) Dipole regime represents the continuation of the dipole–exchange modes at large sample volumes. Among the modes undergoing multiple anti-crossings, we show only the Kittel mode, indicated by the upper green dash-dotted line.
(d) Representative mode profiles. Spatial distributions of $|\mathbf m(\mathbf r)|^2$ together with the dynamical magnetization flow (red arrows) are presented for the lowest-energy exchange modes shown in panel a). In all panels, the external field is $H^{\rm{ext}} = 5.5 \cdot 10^4 \si{A/m} =0.7\cdot 10^3 \si{Oe}$ and is oriented along the $\mathbf{z}$ axis.}
\end{figure*} 

The key distinction between a finite cylinder and a sphere is twofold.
First, a cylinder shape lacks full rotational symmetry: only continuous rotations about the cylinder axis remain a symmetry of the sample. Second, even for a spatially uniform equilibrium magnetization $\mathbf M_0 = M_s \hat{\mathbf z}$, the static demagnetizing field $\mathbf H^{\mathrm{dm}}_0(\mathbf r)$ inside a 
finite cylinder is generally \emph{non-uniform}. This non-uniformity stems from the presence of planar end faces, 
which produce a spatially varying internal
magnetostatic field. Both features play an essential role in the structure and degeneracies of spin-wave modes in cylindrical resonators. In this section, we step-by-step disentangle the roles of exchange, non-uniform static demagnetization $\mathbf{H}^{\rm{dm}}_0$, and non-local dipolar interaction. The case of the non-uniform static magnetization $\mathbf{M}_0$ is discussed in Sec.\ref{sec:EquilibriumMagnetization}. 

\subsection{Exchange modes with homogeneous static field}
\label{sec:cyl-exchange}

The static demagnetizing field  given by Eq.~\ref{eq:dm-integral-kernel} can be decomposed  into a homogeneous part and a spatially varying contribution, shown in Fig.~\ref{fig:Cylinder}
\begin{equation*}
\mathbf H^{\mathrm{dm}}_0(\mathbf r)
=
\overline{\mathbf H}^{\mathrm{dm}}_0
+
\delta\mathbf H^{\mathrm{dm}}_0(\mathbf r),
\qquad
\overline{\mathbf H}^{\mathrm{dm}}_0
=
- N_z M_s \hat{\mathbf z}.
\label{eq:Hdm0_split}
\end{equation*}
The spatially varying part $\delta\mathbf H^{\mathrm{dm}}_0(\mathbf r)$ acts as a geometry-induced effective potential for spin waves and is present even for a uniform equilibrium magnetization, which can be achieved in a cylinder for sufficiently
large bias fields (see discussion in Sec.\ref{sec:EquilibriumMagnetization}).

At the first step, we neglect the spatially varying part of the demagnetization field and replace
$\mathbf H^{\mathrm{dm}}_0(\mathbf r)\rightarrow
\overline{\mathbf H}^{\mathrm{dm}}_0$. The total homogeneous static field is
\begin{equation*}
\overline{\mathbf H}_0^{\text{eff}}
=
\mathbf H^{\mathrm{ext}}+\overline{\mathbf H}^{\mathrm{dm}}_0
=
\big(H^{\mathrm{ext}}-N_z M_s\big)\hat{\mathbf z}.
\end{equation*}

The exchange-only equation then reads
\begin{equation}
-\partial_t \mathbf m
=
\gamma\Big[
\ell_{\mathrm{exch}}^2\,\mathbf M_0\times\nabla^2\mathbf m
-
\overline{\mathbf H}_0^{\text{eff}}\times\mathbf m
\Big],
\label{eq:cyl_exch_cart}
\end{equation}
which can be reduced to a governing equation identical to the  case of a sphere (see Eq.~\eqref{eq:sphere-exchange-governing-equation})
\begin{equation*}
-i\partial_t \mathbf m
=
\gamma\,\hat S_z
\big(
\ell_{\mathrm{exch}}^2\Delta-\overline{H}_0^{\text{eff}}
\big)\mathbf m =\mathcal H^{\rm{uni}}_{\rm{exch}}\mathbf m,
\label{eq:cyl_exch_operator}
\end{equation*}
where $\Delta$ is the Laplacian, defined as
\begin{equation*}
    \Delta=\frac{1}{\rho}\partial_{\rho}\rho\partial_{\rho} +\partial_z^2-\frac{\hat{L}_z^2}{\rho^2}, \qquad \hat{L}_z = -\textit{i}\partial_{\varphi}. 
\end{equation*}

Within the uniform-field exchange approximation, the effective Hamiltonian $\mathcal H^{\rm{uni}}_{\rm{exch}}$ possesses a well-defined set of spatial and spin symmetries, which determine the structure and degeneracies of the spectrum.

First, as in the spherical exchange Hamiltonian, $\mathcal H^{\rm{uni}}_{\rm{exch}}$ is also commutes with the spin and orbital operators $\hat {S}_z$ and $\hat {L}_z$. Since these operators were defined in Section~\ref{sec:sphere-exch-model}, it follows that the Hamiltonian commutes with the total angular momentum projection $\hat {J}_z$ (see Appendix~\ref{appx:commute-with-J}):
\begin{equation}
[\mathcal{H_{\mathrm{exch}}^{\mathrm{uni}}},\, \hat {J}_z]=0,
\end{equation}
and therefore the quantum numbers $j_z$, $l_z$, and polarization are conserved in the exchange dominated regime under consideration.

Second, for a cylinder centered at $z=0$ with identical boundary conditions at both top and bottom faces, the system is invariant under reflection with respect to the midplane, $z \rightarrow -z$. For an axial-vector magnetization field, this symmetry leads to a conserved total parity operator $\widehat P_z$. Thus, $[\mathcal H^{\rm{uni}}_{\rm{exch}}, \hat P_z] = 0 $ and  eigenvalues $P_z=\pm 1$ characterize  modes that are even or odd with respect to the midplane reflections. 

Finally, in the considered  approximation the Hamiltonian is invariant under coordinate inversions in the transverse plane,
$\mathcal I_x:(x,y,z)\rightarrow(-x,y,z)$ and
$\mathcal I_y:(x,y,z)\rightarrow(x,-y,z)$. These transformations leave the scalar Laplacian and the homogeneous static field unchanged, while mapping the azimuthal angle as $\varphi \rightarrow -\varphi$.
As a consequence, they relate eigenmodes with opposite orbital angular momentum projections $l_z$ and $-l_z$ and imply an exact degeneracy of the exchange spectrum with respect to the sign of $l_z$. This degeneracy will be lifted  once symmetry-breaking contributions, such as the non-local dipolar interaction, are taken into account.

Passing to circular components $m_\pm$ Eq.~\eqref{eq:transform-cart-circ1} diagonalizes $\hat S_z$. For harmonic modes $m_\pm(\mathbf r)e^{-i\Omega t}$ one obtains
\begin{equation}
\Omega^{\rm{uni}} m_\pm
=
\pm\gamma\big(\ell_{\mathrm{exch}}^2 M_s\Delta-\overline {H}_0^{\text{eff}}\big)m_\pm .
\label{eq:cyl_pm}
\end{equation}

The exchange boundary conditions follow from
$\mathbf n~\cdot~\nabla\mathbf m|_{\partial V}=0$ and reads
\begin{equation}
\left.\partial_\rho m_\pm\right|_{\rho=r_0}=0,
\qquad
\left.\partial_z m_\pm\right|_{z=\pm h_0/2}=0 .
\label{eq:cyl_bc}
\end{equation}

The eigenfunctions factorize as
\begin{equation}
m_\pm^{(l_z, n ,p)} (\rho,\varphi, z)
=
C_{l_z, n, p}\,
J_{l_z}\!\left(\kappa_{p,l_z}\frac{\rho}{r_0}\right)
Z_n(z)\,e^{i l_z\varphi},
\end{equation}

where the normalization $\mathcal{C}^{-1}_{l_z,n}=\int_V{|m_{\pm,(l_z, n)}|^2 \ dV}$,  $J_{l_z}(\rho)$ is a Bessel function, and $Z_n(z)$ is the standing-wave
\begin{equation}
Z_n(z)=
\begin{cases}
\cos\!\left(\dfrac{\pi n z}{h_0}\right), & P_z=+1,\\[6pt]
\sin\!\left(\dfrac{\pi n z}{h_0}\right), & P_z=-1,
\end{cases}
\end{equation}
$n$ is the axial quantum number, and $\kappa_{p,l_z}$ is the $p$-th root (numbering starts at zero) of $J'_{l_z}(x)=0$.

The corresponding exchange eigenfrequencies are
\begin{equation}
\Omega^{(0)}_{l_z, n, p}
=
\gamma\overline{H}_0^{\text{eff}}
+
\Omega_M\ell_{\mathrm{exch}}^2
\left[
\left(\frac{\kappa_{p,l_z}}{r_0}\right)^2
+
\left(\frac{\pi n}{h_0}\right)^2
\right],
\label{eq:cyl_exch_freq}
\end{equation}
where $\Omega_M = \gamma M_s$. These eigenfrequencies are plotted in Fig.~\ref{fig:Cylinder} (a) for different quantum numbers $l_z$ and $n$. The corresponding mode profiles are shown in Fig.~\ref{fig:Cylinder}(c).

Equation~\eqref{eq:cyl_exch_freq} shows that all non-uniform exchange modes scale inversely with respect to the radius (height) of the cylinder at the fixed aspect ratio $2r_0/h_0=1$. Thus, upon increasing the cylinder size, all exchange branches collapse towards the uniform precession frequency $\gamma\overline{H}_0^{\text{eff}}$, while for nanoscale radii, the spectrum is dominated by the $1/r_0^2$ exchange scaling.  The only exception is the uniform Kittel-like $ l_z=0, n=0, p=0$ mode, which has no dispersion and no size dependence.  We also see that all the modes are degenerate with respect to the $l_z$ number.

\subsection{ Effect of the non-uniform static demagnetizing field}
\label{sec:cylinder-nonuni}

{We now retain $\delta\mathbf H^{\mathrm{dm}}_0(\mathbf r)$ while still neglecting
the non-local dipolar interaction.
Decomposing the torque
$\mathbf H^{\mathrm{dm}}_0\times\mathbf m$ in the circular basis, one finds that
only the longitudinal component $\delta H^{\mathrm{dm}}_{0,z}(\mathbf r)$
contributes to the linearized transverse dynamics. Indeed, the non-uniform static demagnetizing field contributes an additional torque term, which  can be written in the circular basis and expressed via spin-1 matrices as follows}
\begin{equation*}
    \label{eq:cylinder-dm-field-torque-term}
    \mathbf{\tilde{H}}^{\rm{dm}}_0\times \mathbf{\tilde{m}}=\Big(H^{\rm{dm}}_{+}\hat{\tilde{S}}_+ +H^{\rm{dm}}_z\hat{\tilde{S}}_z+ H^{\rm{dm}}_-\hat{\tilde{S}}_- \Big)\cdot\mathbf{\tilde{m}},
\end{equation*}

{thus this term acquires $0$ -- spin component, that enters the system Eq.~\eqref{eq:cyl_exch_cart} as the third equation}
\begin{equation*}
    H^{\rm{dm}}_{+}m_- -H^{\rm{dm}}_{-}m_+ = 0.
\end{equation*}
{This torque component, however, generate terms proportional to $m_z$ and therefore do
not affect the dynamics in the linear approximation with $\mathbf m\perp\mathbf
M_0$.}

As a result, the non-uniform static field enters as a local operator
\begin{equation}
\delta\mathcal {H}_{\rm{exch}}(\mathbf r)
=
-\gamma\,\hat S_z\,\delta H^{\mathrm{dm}}_{0,z}(\mathbf r).
\label{eq:U_def}
\end{equation}

Importantly, although $\mathcal U(\mathbf r)$ is spatially dependent, it preserves
axial symmetry and coordinate inversions,
\[
[\delta\mathcal{H}_{\rm{exch}},\hat J_z]=[\delta\mathcal{H}_{\rm{exch}},\mathcal{I}_x]=[\delta\mathcal{H}_{\rm{exch}},\mathcal{I}_y]=0,
\]
\[
[\delta\mathcal{H}_{\rm{exch}},\hat P_z]=0.
\]
Therefore, the spatial non-uniformity of the static demagnetizing field alone
\emph{does not lift} the $|l_z|$ degeneracy of the exchange spectrum.
Its effect is instead to destroy separability in $(\rho,z)$ and to mix modes
with different axial and radial indices $(n,p)$ within a fixed $(j_z,P_z)$
sector.

\subsection{Dipole-exchange modes}
\label{sec:cyl-dip-exchange}

Finally, we restore the non-local dipolar term.
The full linear operator then reads
\begin{equation}
\mathcal{H}_{\text{dip-exch}} =  \mathcal{H}_{\rm{exch}}^{\rm{uni}} + \delta\mathcal{H}_{\rm{exch}}+\mathcal{H}_{\rm{dip}}.
\end{equation}

The numerically calculated dipole-exchange spectra are presented in Fig.~\ref{fig:Cylinder}(b). The eigen-branches with same $|l_z|$, that were initially degenerate in the exchange regime, split in the dipole-exchange regime. This behavior is due to symmetry breaking, leading to the loss of conservation of ${\mathcal{I}_x}$, and ${\mathcal{I}_y}$ parities, and states with the opposite $l_z$ are now not connected by $\varphi\to-\varphi$ transform. This symmetry is lost because of the dipole kernel isn't commute with coordinate inversion operator (see Appendix~\ref{appx:inversion-symmetries}). This degeneracy lifting explanation improves on earlier interpretations \cite{lim2021ferromagnetic}, that correctly attributed the splitting to the loss of coordinate inversion symmetry, but related this loss to equilibrium magnetization $\mathbf{M}_0 \parallel\mathbf{\hat{z}}$ and external magnetic field $\mathbf{H}^{\rm{ext}}\parallel\mathbf{\hat{z}}$  breaking time-reversal. However, both $\mathbf{M}_0$ and $\mathbf{H}^{\rm{ext}}$ are already present in the purely exchange problem, where $\pm l_z$ degeneracy can still persist. Our symmetry analysis demonstrates that the reduced exchange problem remains invariant under coordinate inversions, but losses it once the self-consistent dipolar term is retained. As a result, there is no longer a symmetry that enforces $\Omega(l_z) =\Omega(-l_z)$ for a fixed biased state, and the $\pm l_z$ degeneracy can be lifted even though axial rotation symmetry is preserved.

Thus, the only invariants that survive in dipole-exchange regime are $\hat{J}_z$ and full middle plane reflection $\hat{\mathcal{P}}_z$. In the spectrum, this manifests itself also as the appearance of avoided crossings between modes belonging to the same
$(j_z,P_z)$ symmetry sector, in direct analogy with the spherical case discussed in Sec.~\ref{sec:sphere-dipole-exchange}.

\subsection{Dipole interaction limit}

{The  spectrum in the  dipole dominated regime is shown in Fig.~\ref{fig:Cylinder}(c), where the mode branches are colored according to the dipole--exchange modes from which they evolve. Dipolar-dominated regime retains the same conserved quantities as the full dipole--exchange system, namely the total parity $\hat{P}_z$ and the $z$-component of the total angular momentum $\hat{J}_z$, thus the dipolar modes are labeled with the same $j_z$ quantum number. 
At a fixed aspect ratio, the frequency of the dipolar mode does not depend on the cylinder volume, which results in the dispersionless eigenlines shown in Fig.~\ref{fig:Cylinder}(c). This behavior is expected for modes governed solely by the internal dipolar field rather than by exchange-induced wavelength quantization.}

{An important difference from the sphere concerns the behavior of the Kittel mode. Unlike in sphere, the cylindrical uniform mode in exchange limit with $(l_z,n,p)=(0,0,0)$, hybridizes with the other modes of the same symmetry sector after entering the dipole--exchange regime, which can be clearly seen in the gray rectangular in  Fig.~\ref{fig:Cylinder}(b). Therefore, it does not remain a spatially uniform mode in the dipole limit and is labeled as $j_z=1$ (green solid line). In contrast, another mode becomes uniform in the dipole limit and is identified as the Kittel mode in Fig.\ref{sec:cmt}. 
}


\begin{figure*}[t]
    \centering
    \includegraphics[width=1\textwidth]{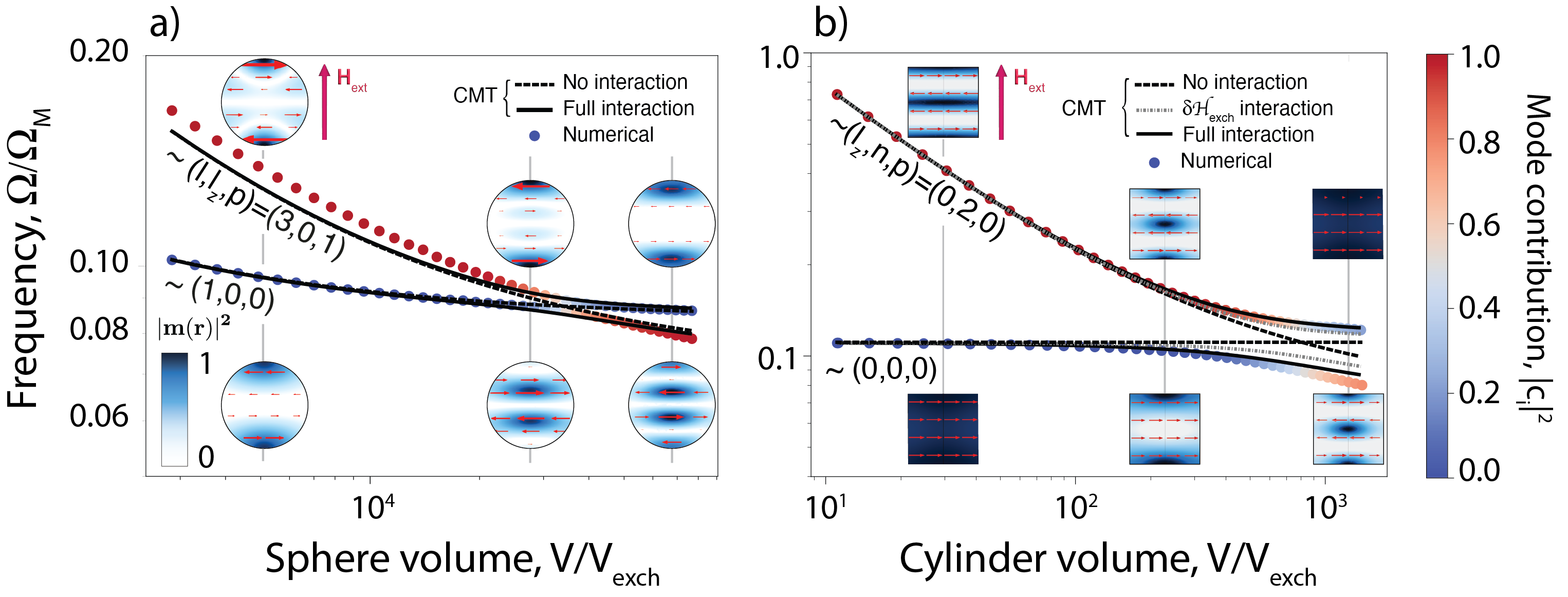}
    \caption{Plots of the spin-wave spectrum (logarithmic axis) calculated using magnonic coupled-mode theory (CMT), compared with the numerical solution of the dipole–exchange problem (filled circles). Crossing (dashed black) lines show the CMT solution with self-energy terms  (non-interacting limit), while anti-crossing (solid black) lines include both self-energy and interaction terms.
  Panel (a) shows dipolar-induced coupling between the spin-wave modes originating from the exchange modes $(1,0,0)$ and $(3,0,1)$ in sphere.
Panel (b) shows coupling between the $(0,0,0)$ and $(0,2,0)$ modes in cylinder; anti-crossing (dashed gray) lines indicate CMT results including only the non-uniform static-field operator $\delta \mathcal{H_{\mathrm{exch}}}$. 
Insets show the evolution of the mode profiles before, near, and after the interaction at the volume marked by vertical gray lines.
\label{fig:CMT}}
\end{figure*}

\section{Coupled Mode Theory for Dipole--Exchange Spin Waves }
\label{sec:cmt}
The analysis above shows that exchange eigenmodes provide a natural and symmetry-adapted basis for describing dipole--exchange spin waves in confined three-dimensional resonators. In this basis, the effects of the non-uniform static demagnetizing field and the non-local dipolar interaction appear as mode-coupling terms that lift degeneracies and produce avoided crossings. This observation motivates a coupled-mode formulation, which we develop in the following section.
A related approach for treating spin-wave dynamics in finite magnets was proposed by Mills \cite{mills2006quantum} in the context of quantum dipole–exchange magnonics. In that formalism the eigenproblem is solved for coupled magnetization and magnetostatic potential fields, which together form a complete set of dipole–exchange eigenfunctions that can be normalized over the entire space and subsequently used for spin-wave quantization. While this framework provides a rigorous continuum description, the explicit use of the scalar potential complicates the analysis and obscures the relative roles of the local exchange and nonlocal dipolar interactions. To elucidate the dynamical crossover between these interaction regimes, we develop a general spin-wave coupled-mode theory (CMT). Following the spirit of the dipole–exchange theory of Slavin and Kalinikos for thin ferromagnetic films \cite{kalinikos1986theory}, we generalize their approach to arbitrary three-dimensional systems and apply it to spherical and cylindrical geometries. This reduces the dipole–exchange problem to a finite coupled-mode system that directly describes mode hybridization in the dipole–exchange regime and provides a transparent connection between the exchange-dominated and dipolar limits.
 
\subsection{Exchange eigenmodes as a basis}
\label{subsec:cmt_basis}

We begin by recalling that in the absence of non-local dipolar interactions and Gilbert damping, the linearized spin-wave dynamics is governed by a local Hermitian operator $\mathcal{H}_{\rm{exch}}$, whose explicit form depends on the resonator geometry
(Sec.~\ref{sec:sphere-exch-model} and \ref{sec:cyl-exchange}). 
The corresponding exchange eigenmodes provide a natural basis for constructing a coupled-mode description
of the full dipole--exchange problem. Let $\{\mathbf m_\nu(\mathbf r)\}$ denote the set of exchange eigenmodes,
satisfying
\begin{equation}
\mathcal H_{\rm{exch}}\,\mathbf m_\nu(\mathbf r)
=
-\Omega^{e}_\nu\,\mathbf m_\nu(\mathbf r),
\label{eq:exch_eigenproblem}
\end{equation}
together with the exchange boundary conditions
$\mathbf n\cdot\nabla\mathbf m_\nu|_{\partial V}=0$
and the transversality constraint
$\mathbf m_\nu\perp\mathbf M_0$.
Here $\nu$ is a collective index labeling the spatial mode numbers:
$\nu=(l,l_z,p)$ for a sphere and $\nu=(l_z,n,p)$ for a finite cylinder.

Because $\mathcal H_{\rm{exch}}$ is Hermitian, the exchange eigenmodes form  orthonormal set with respect to the volume inner product
\begin{equation}
\langle \mathbf m_\nu , \mathbf m_{\nu'} \rangle
\equiv
\frac{1}{V}
\int_V d^3 \mathbf r\,
\mathbf m_\nu^\dagger(\mathbf r)\cdot\mathbf m_{\nu'}(\mathbf r)
=
\delta_{\nu\nu'} .
\label{eq:exch_orthonorm}
\end{equation}
With this normalization, the set $\{\mathbf m_\nu\}$ forms a complete basis for expanding any admissible transverse dynamic magnetization field inside the resonator.

In the coupled-mode formulation, the full dynamic magnetization is expanded over the exchange eigenmodes,
\begin{equation}
\mathbf m(\mathbf r,t)
=
\sum_\nu c_\nu(t)\,\mathbf m_\nu(\mathbf r),
\label{eq:mode_expansion}
\end{equation}
The time-dependent coefficients $c_\nu(t)$  describe the expansion of spin waves in the exchange basis.
Inserting this expansion into the full linearized equation of motion and projecting onto the exchange basis leads to a closed set of coupled equations for the amplitudes $c_\nu(t)$, which we derive in the following subsection.

\subsection{ General formulation of the dipole--exchange equation}
\label{subsec:cmt:b}
Our starting point is the linearized equation of motion describing the dynamic magnetization,
\begin{equation}
-i\partial_t \mathbf m(\mathbf r,t)
=
\big(
\mathcal H^{\rm{uni}}_{\rm{exch}} + \delta\mathcal{H}_{\rm{exch}}(\mathbf r) + \mathcal{H}_{\rm{dip}}\big)\mathbf m(\mathbf r,t),
\label{eq:full_operator_eq}
\end{equation}
where $\mathcal H^{\rm{uni}}_{\rm{exch}}$ is the local exchange operator with a homogeneous static
field, $\delta\mathcal{H}_{\rm{exch}}(\mathbf r)$ is the local operator originating from the spatially non-uniform static demagnetizing field, and $\mathcal{H}_{\rm{dip}}$ denotes the non-local dipolar interaction operator.

Using the mode expansion Eq.~\eqref{eq:mode_expansion} 
and substituting it into Eq.~\eqref{eq:full_operator_eq}, we project the
equation of motion onto the exchange basis by taking the inner product Eq.~\eqref{eq:exch_orthonorm} with
$\mathbf m_\mu^\dagger(\mathbf r)$.
This yields a closed system of coupled equations for the mode amplitudes
$c_\nu(t)$,
\begin{equation}
i\,\dot c_\mu(t)
=
\Omega^{e}_\mu\,c_\mu(t)
-
\sum_\nu
\big(
U_{\mu\nu}
+
K_{\mu\nu}
\big)c_\nu(t).
\label{eq:coupled_equations}
\end{equation}

Here the matrix elements of the local operator $\delta\mathcal{H}_{\rm{exch}}(\mathbf r) $ are given by
\begin{equation}
U_{\mu\nu}
=
\frac{1}{V}
\int_V d^3 r\,
\mathbf m_\mu^\dagger(\mathbf r)\cdot
\delta\mathcal{H}_{\rm{exch}}(\mathbf r)\cdot\mathbf m_\nu(\mathbf r),
\label{eq:U_matrix}
\end{equation}
while the matrix elements of the non-local dipolar operator $\hat K$ read
\begin{equation}
K_{\mu\nu}
=
\frac{\gamma M_s}{V}
\int_V d^3 r
\int_V d^3 r'\,
\mathbf m_\mu^\dagger(\mathbf r)\cdot\hat{S}_z\cdot
\hat K(\mathbf r-\mathbf r')\mathbf m_\nu(\mathbf r').
\label{eq:K_matrix}
\end{equation}

Eq.~\eqref{eq:coupled_equations} represents the most general linear dipole--exchange problem in a confined resonator formulated in the exchange mode basis. It can be reformulated in the frequency domain by substituting $c_{\mu}(t)\propto e^{-i\Omega t}$, where $\Omega$ will constitute the eigen frequencies of modes and  Eq.~\eqref{eq:full_operator_eq}  will acquire a form of  algebraic systems. 
{In practical calculations, Eq.~\eqref{eq:coupled_equations} is evaluated in a finite, symmetry-adapted exchange-mode basis.
Within a fixed $(j_z,P_z)$ sector, we order the positive-frequency exchange modes by increasing exchange frequency and use the first $N$ modes to form the nested Hamiltonian $H_N$.
The convergence of this truncation is documented in Appendix~\ref{app:cmt-stability} through local and cumulative relative-RMS changes of the low-lying eigenfrequencies as the exchange-mode basis is enlarged.
Importantly, the eigenvalues of a fixed $N$-mode problem are invariant under a permutation of its basis functions: a reordering gives the similar matrix $P^{\dagger}H_NP$.
The ordering only specifies the path by which the infinite basis is truncated; throughout we use the deterministic exchange-frequency ordering stated above.
}
The matrix of this system consists of diagonal terms, which represent the self-energies, and off-diagonal elements, which describe mode hybridization. It is determined by the non-uniform static demagnetizing field $U_{\mu\nu}$ and by the non-local dipolar interaction $K_{\mu\nu}$.
It is important to emphasize the qualitative difference between these two coupling mechanisms. The operator $\delta\mathcal{H}_{\rm{exch}}(\mathbf r)$ is local in space and originates solely from the spatial variation of the static demagnetizing field.
As discussed in Sec.~\ref{sec:cylinder-nonuni}, this term preserves all symmetries of the exchange Hamiltonian and therefore cannot lift symmetry-protected degeneracies by itself. 
In contrast, the non-local operator $\hat K$ couples magnetization dynamics at different spatial points and breaks the coordinate inversion symmetries responsible for the degeneracy of modes (Sec.~\ref{appx:inversion-symmetries}). As well, it preserves axial symmetry with respect to the magnetization direction and therefore commutes with the total angular momentum projection,
$[\mathcal{H}_{\rm{dip}},\hat J_z]=0$. As a result, dipolar coupling is allowed only between exchange modes with the same value of $j_z$. Then, modes belonging to different $j_z$ sectors remain symmetry-protected. As we mentioned above, the exchange modes obey a rigid constraint between $j_z$ and $l_z$, namely $j_z=l_z+1$. Consequently, modes with different $l_z$ cannot couple.
Finally, we note that in the absence of Gilbert damping the operators $\mathcal H^{\rm{uni}}_{\rm{exch}}$, $\delta\mathcal{H}_{\rm{exch}}$, and $\mathcal{H}_{\rm{dip}}$ are Hermitian with respect to the inner product defined in Eq.~\eqref{eq:exch_orthonorm}. As a result, the matrix
appearing on the right-hand side of Eq.~\eqref{eq:coupled_equations} is Hermitian,
and the resulting dipole--exchange eigenfrequencies are strictly real.

\subsection{ Coupled mode formalism in a sphere}
\label{subsec:cmt:c}

We now apply the general dipole--exchange formulation developed above to the case of a uniformly magnetized sphere.
Although the spin-wave spectrum of a sphere has been extensively studied, this geometry provides a particularly transparent example for illustrating how the coupled-mode formulation captures symmetry-protected degeneracies, their lifting, and the emergence of avoided crossings.

{First of all}, for a uniformly magnetized sphere
$\mathbf M_0 \parallel \hat{\mathbf z}$, the operator $\delta\mathcal{H}_{\rm{exch}}$ vanishes since the demagnetization field is homogeneous in sphere. 

For one $j_z$ sector, the coupled-mode equations
\eqref{eq:coupled_equations} reduce to a matrix eigenvalue problem of the form
\begin{equation}
\Omega\,c_{l, l_z, p}
=
\Omega^{e}_{l,l_z, p}\,c_{l,l_z,p}
-
\sum_{l', p'}
K^{l l_z p}_{l' l_z p'}\,c_{l' l_z p'},
\label{eq:sphere_cmt_matrix}
\end{equation}
where the indices $l'$, and $p'$ run over all possible values and $l_z=j_z-1$. 

To illustrate the method, we begin by applying the CMT framework to two representative cases. 
First, the uniform mode  in a sphere, characterized by $(l,l_z,p)=(0,0,0)$, has a particular clarity in this picture. The uniform mode generates a uniform demagnetizing field and is therefore an eigenfunction of $\mathcal{H}_{\rm{dip}}$,
\begin{equation}
\label{eq:kittel-mode-dipolar-action}
\mathcal{H}_{\rm{dip}}\mathbf m_{000} = -\frac{1}{3} \gamma M_s \mathbf m_{000},
\end{equation}
\roundtwo{so its off-diagonal coupling matrix elements with all other orthogonal exchange modes vanish, including modes in the same $(j_z,P_z)$ symmetry sector. Its dipolar self-energy shift is independent of the sphere volume, and the Kittel branch therefore remains dispersionless at fixed external field. The crossing with the branch originating from $(l,l_z,p)=(2,0,1)$ in Fig.~\ref{fig:Sphere}(b) explicitly illustrates this decoupling: the two branches cross without an avoided crossing despite having identical symmetry labels. Thus, belonging to the same symmetry sector is necessary, but not sufficient, for hybridization.} 

As another example, we have investigated the coupling between two spin-wave modes, originating from the exchange modes $(1,0,0)$ and $(3,0,1)$, which emerges when dipolar interactions are included. These modes belong to the same domain with $j_z=1$. Fig. \ref{fig:CMT} (a) shows the appearance of anticrossing feature in the spectrum with the change of the particle volume.  The solid black lines correspond to the CMT solution including both the dipolar self-energy (diagonal matrix elements of $\mathcal{H}_{\rm{dip}}$) and the off-diagonal coupling between the two exchange modes. In contrast, the dashed black lines show the result obtained when mode coupling is artificially suppressed, retaining only the diagonal self-energy terms.The numerically calculated eigenfrequencies obtained by directly solving the dipole–exchange problem are shown as filled circles. The color of each circle indicates the normalized magnitude of the associated eigenvector of the system matrix given in Eq.~\eqref{eq:sphere_cmt_matrix}, thereby demonstrating a continuous exchange of modal character across the interaction region. The same mode flipping can also be observed in six representative mode profiles showing the  dynamical magnetization direction with red arrows.

At small volumes (exchange-dominated regime), the two branches follow the uncoupled exchange dispersion and remain well separated due to self-energy terms. As the volume increases, the frequencies of the two exchange modes approach each other and would cross in the absence of coupling (only diagonal self-energy terms).  However, once the off-diagonal dipolar matrix elements are included, this crossing is replaced by a pronounced avoided crossing. The minimal frequency splitting at the anticrossing is fully captured by the coupled-mode theory and is in excellent agreement with the numerical results. 

\subsection{Coupled mode formalism in a cylinder}
\label{subsec:cmt:d}

We now apply the coupled-mode theory to the case of a uniformly magnetized finite cylinder with equilibrium magnetization
$\mathbf M_0 \parallel \hat{\mathbf z}$. First, unlike in a sphere, the static demagnetizing field inside a uniformly magnetized cylinder is spatially non-uniform.
Consequently, the local operator $\delta\mathcal{H}_{\rm{exch}}(\mathbf r)$ does not vanish and contributes to the coupled-mode equations.
As discussed in Sec.~\ref{sec:cyl-dip-exchange}, the mode coupling can appear only    between two spin-wave modes belong to the same $(j_z, P_z)$ symmetry sector. 

For the cylindrical geometry, the coupled-mode equations
\eqref{eq:coupled_equations} take the form
\begin{equation}
\Omega\,c_{l_z, n, p}
=
\Omega^{e}_{l_z, n, p}\,c_{l_z, n, p}
-
\sum_{n', p'}
\left(
U^{l_z n p}_{l_z n' p'}
+
K^{l_z n p}_{l_z n' p'}
\right)
c_{l_z n' p'} ,
\label{eq:cyl_cmt_matrix}
\end{equation}
where the indices $l_z$, $n$,  and $p$ label the azimuthal, z-axial and radial exchange quantum numbers, respectively. As in the spherical case, the matrix can be brought to a block-diagonal form,
with each block corresponding to a fixed value of the conserved quantum numbers $j_z$ and $P_z$. The diagonal matrix elements describe self-energy shifts due to both the static field inhomogeneity and the dipolar interaction, while the off-diagonal elements account for mode hybridization.

A notable difference with respect to spherical geometry concerns even the lowest (Kittel) mode in a finite cylinder.
Although this mode remains nearly spatially uniform for sufficiently small aspect ratios (see Fig.~\ref{fig:Cylinder} b), dipolar interaction couples this mode to the other modes as the Kittel mode is no longer an exact eigenfunction of the non-local dipolar operator $\mathcal{H}_{\rm{dip}}$.  Figure~\ref{fig:CMT} (b) shows this hybridization behavior of Kittel mode $(l_z, n,p)=(0,0,0)$ mixing with  $(0,2,0)$ mode, which belong to the same symmetry sector $j_z=1, P_z=1$. 

The solid black lines demonstrate the full coupled-mode solution, including both $\delta \mathcal{H_{\mathrm{exch}}}$ and $\mathcal{H_{\mathrm{dip}}}$. The dashed black lines correspond to a calculation where mode coupling is artificially removed, leaving only the diagonal self-energy terms. The dashed gray lines represent the case where the diagonal self-energy is included together with only the off-diagonal components of $\delta \mathcal{H_{\mathrm{exch}}}$. 
From the graph, it is evident that the non-uniform static demagnetizing field also gives rise to mode coupling. 
The numerical results obtained from the direct dipole--exchange calculations and shown as filled circles are in close agreement with the solid black lines.
{For the calculation in Fig.~\ref{fig:CMT}(a), the sums over $l'$ and $p'$ in Eq.~\eqref{eq:sphere_cmt_matrix} were restricted to the two interacting exchange modes $(l,l_z,p)=(1,0,0)$ and $(3,0,1)$; the cylindrical calculation in Fig.~\ref{fig:CMT}(b) was analogously restricted to $(l_z,n,p)=(0,0,0)$ and $(0,2,0)$.
This two-mode truncation retains their mutual hybridization and diagonal self-energy shifts but omits the additional renormalization produced by other exchange modes in the same symmetry sector.
Being away from the avoided crossing suppresses the mutual mixing of the two selected modes, but it does not remove these weak contributions from the omitted modes.
The effect is most visible for the higher-order spherical mode $(3,0,1)$, whose more structured spatial profile permits appreciable overlap with additional exchange modes and is therefore represented less completely by the two-mode subspace.
The lower-order $(1,0,0)$ mode has a simpler spatial structure and is already represented accurately by this truncation, so no comparable offset is resolved.
The same truncation effect is present for the cylinder, but the residual discrepancy is smaller because both interacting cylinder modes are of lower order and are better described within the selected two-mode subspace.}
The numerical results, obtained from the direct dipole–exchange calculations, shown as filled circles, coincide with the solid black lines. This agreement confirms the correctness of our coupled-mode theoretical construction.

\section{Discussion and outlook}
\label{sec:Discussion}
Within this section, we would like to provide additional discussion of the obtained results and offer a further outlook for the potential development of the work. 
\subsection*{Equilibrium magnetization states}
\label{sec:EquilibriumMagnetization}
The equilibrium magnetization state of a system plays a leading role in defining the spin-wave mode structure.
Throughout this article, we assumed a uniform equilibrium magnetization profile in a ferrimagnetic sphere and cylinder to simplify the equations. However, that may not generally be the case, especially for micron-sized structures. This discussion aims to shed more light on the bounds of applicability of our assumptions. 
\begin{figure}
\label{fig:Sergey}
    \centering
    \includegraphics[width=\columnwidth]{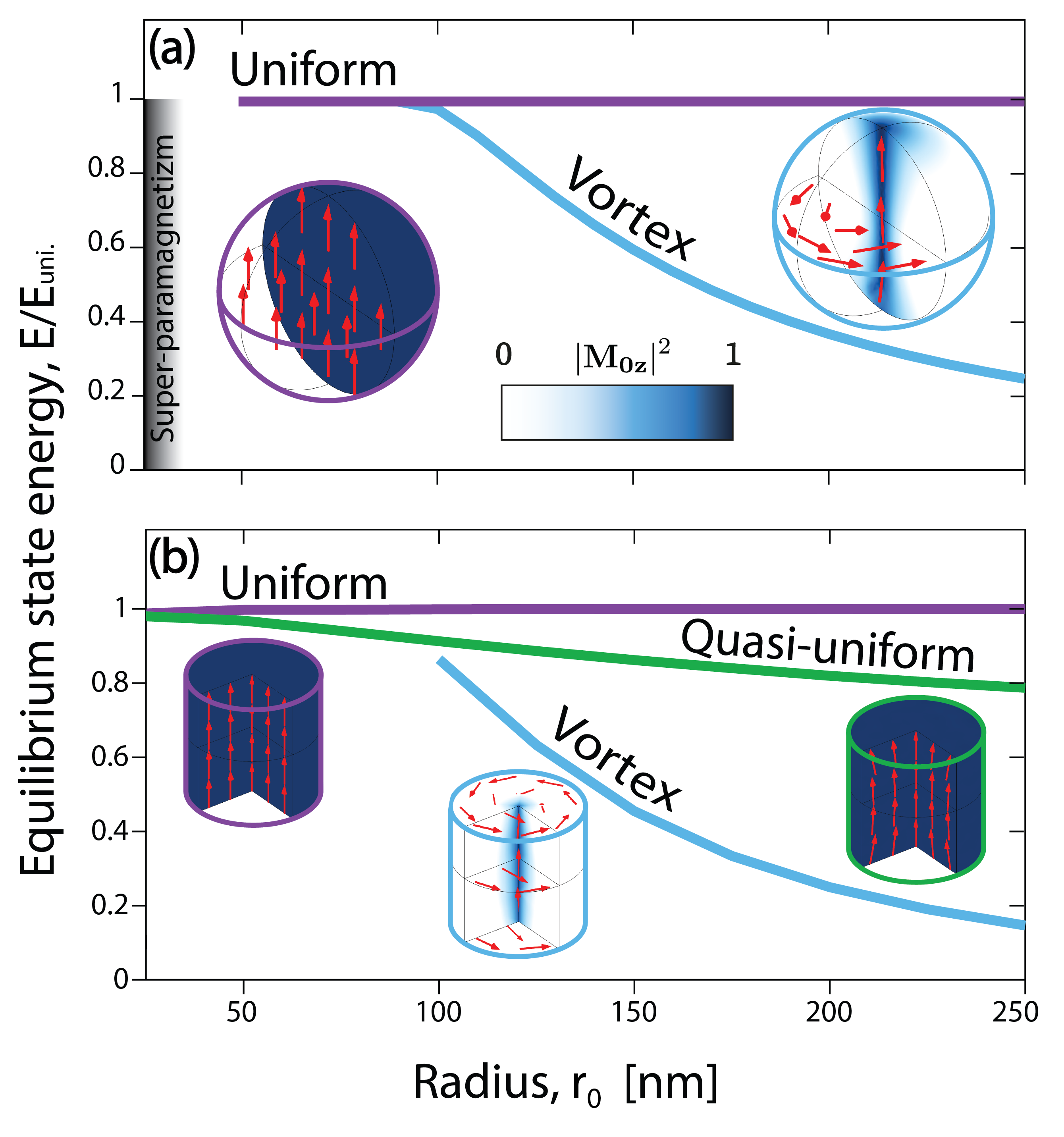}
    \caption{\label{fig:Figure4}
   In zero external field, magnetic state energy $E$ is shown as a function of the sample radius. The uniformly magnetized state is indicated as the purple line with energy $E_{\text{uni.}}$. (a) For a sphere, the vortex state (blue line) forms at radii above the critical value $r_{sph} = 90~\si{\nano\meter}$ and has lower energy than the uniform state (purple line).  \label{fig:StaticM0}
   (b) For a cylinder, the unstable uniform state (purple line) transforms into a more energetically favorable quasi-uniform state (green line). The vortex state (blue line), with its core oriented along the cylinder axis, attains the lowest energy for cylinder radii above $r_{cyl} = 100~\si{\nano\meter}$ .
    }
\end{figure}
The equilibrium configuration of the magnetization can be obtained by minimizing the free-energy functional, which depends on the external field, the demagnetizing field, the exchange field, and the magnetization itself \cite{rameshti2022cavity}. 
\begin{equation}
    \begin{split}
    E = \int_V \mathrm{d}^3 r \bigg[& \frac{\mu_0 A}{M_s} \sum_{i = x,y,z} \left| \nabla M_{0i} \right|^2 \\
    &-\mu_0\Big(\mathbf{H}^{\text{ext}}+\frac{1}{2}\mathbf{H}^{\text{dm}}\Big)\cdot\mathbf{M}_0\bigg]
    \end{split}
    \label{eq:placeholder_label}
\end{equation}
From energetic point of view, Eq.~\ref{eq:equilibrium-condition} can be written as:
\begin{equation}
    \mathbf{M}_0\times\frac{\delta E}{\delta \mathbf{M}_0} = 0
\end{equation}
This leads to a complicated energy landscape with several local minima, each associated with a distinct equilibrium magnetization state.  
For illustrating that we have done micromagnetic modeling simulations with software NMAG \cite{fischbacher2007systematic} in its \textit{relaxation} mode to calculate equilibrium magnetization states
in the absence of external field, presented in a Fig~\ref{fig:Figure4}. We investigated YIG with parameters given in Sec.~\ref{sec:computation-details}, setting the external field to zero and the temperature to $T=298 K$. Our study focused on ferrimagnetic spheres and cylinders with radii between $r = 50~\si{\nano\meter}$ and $r = 250~\si{\nano\meter}$. 
In micromagnetic calculations we set minimal element sizes of $4~\si{\nano\meter}$ for $r_{sph} = (50-90)~\si{\nano\meter}$ and $6~\si{\nano\meter}$ for $r_{sph} = (100-250
)~\si{\nano\meter}$. 
Meshes were generated with Netgen/NGSolve software.\\

The results of the simulations are shown in Fig.~\ref{fig:Figure4}(a) for the geometry of the sphere. One can see that for the given size parameters there are two equilibrium states: first having commonly known homogeneous magnetization, and second having vortex-type states. This phenomenon is well-known  in micromagnetics physics \cite{usov2009micromagnetics}. For spherical particles with radii below $r_{sph}=50~\si{\nano\meter}$, superparamagnetic behavior appears at a certain point \cite{krishnan2006nanomagnetism}; consequently, all the data shown in Figs.~\ref{fig:Sphere}, \ref{fig:Cylinder} refer to radii larger than this threshold. For the case of cylinder Fig.~\ref{fig:Figure4} (b) with the same aspect ratio $2r_0/h_0 =1$ as in previous section  there appears a third states in the considered range of parameters which has quasi-uniform magnetization distribution and sometimes attributed to as "flower" state \cite{kronmuller2003micromagnetism,usov2009micromagnetics}. Local magnetic moments of such state tilted statically near body boundary forming symmetric flower-like magnetization distribution. As is evident from Fig.~\ref{fig:Figure4} (b), the energy difference between the unstable homogeneous state and the stable quasi-homogeneous state is small; therefore, to simplify the calculations, we have employed the homogeneous state throughout all the sections above.

It can be observed that homogeneous states possess the lowest energy only for structures whose dimensions are sufficiently small, while for sizes exceeding a certain critical value, non-homogeneous magnetization configurations become energetically preferable. Therefore, to achieve a homogeneous magnetization, a sufficiently strong external magnetic field must be applied. Our estimates show that, for the parameter range under consideration, a critical external field of $\mathbf{H^\text{ext}} = 4 \cdot 10^4~ \si{A/m}= 502~\text{Oe}$ is required to obtain a uniform magnetization state in a sphere with radius $r_{sph}=250~\si{\nano\meter}$. In Sec.~\ref{sec:sphere-exch-model} and \ref{sec:computation-details}, we assumed the external field to be several times larger than this critical value, which justifies the use of the homogeneous magnetization approximation.

For further studies, note that when calculating spin waves on a nonuniform stationary magnetization, additional symmetry breaking occurs. Consequently, $j_z$ may cease to be a good quantum number when the equilibrium magnetization texture is not axially uniform, as in the vortex state, because the operator $\widehat {J}_z$ generally fails to commute with the effective exchange Hamiltonian $\mathcal{H_{\mathrm{exch}}}$.

\subsection*{Coupled mode formalism}
We would like to stress that the proposed coupled-mode theory formalism can be extended to other geometries and cases where the analytical or numerical solution of the {\it purely exchange} problem can be easily and quickly obtained. Then, the full exchange-dipole spectrum, which requires much more resources to solve due to integro-differential character of the problem,  can be constructed using semianalytical CMT approach.  Initially proposed by Slavin and Kalinikos for thin film geometry, it can be effectively applied for other finite 3D geometries. One of the straight forward application of this approach would be constructing the spectra of finite structure in the case of the presence of anisotropy \cite{PhysRevB.65.024414}, additional interaction terms such as Dzyaloshinskii-Moriya \cite{PhysRevB.88.184422,PhysRevLett.114.247206,PhysRevB.95.094414}, or external magnetization along the low-symmetry directions\cite{PhysRevB.80.224403,10.1063/1.4907877}. 

The CMT approach will be very helpful  for further analysis of optomagnonic and acoustomagnoninc cross domain coupling in three-dimensional resonators, where the symmetry of optical  \cite{Gladyshev2020, poleva_multipolar_2023, Frizyuk2021} and acoustic \cite{tsimokha_acoustic_2022,deriy_bound_2022} modes will provide particular angular momentum selection rules.

\subsection*{Spin matrices formalism and elliptical polarization.}
The introduced spin-matrices formalism, though well-known in theoretical physics, have not been widely used in the   field of micromagnonics, but, as we believe, can be effectively utilized and extended for many cases. As an example, it can predict the formation of elliptic polarization states      \cite{Farle1998,kalinikos1986theory}. In the circular polarization representation the dipolar term acting on a spin-wave state $\mathbf{m}=(m_+,0,m_-)$, can be rewritten in terms of spin-1 matrices Eq.~\eqref{eq:hel-spin-martices}, and separates into diagonal and spin-mixing terms 
\begin{equation}
\label{eq:dipole-in-circular}
\mathcal{H}_{\rm{dip}}=\gamma M_s\int_V{d^3\mathbf{r}' \ \Big(k\hat{S}_z^2+\textit{i}K_{xy}\big[\hat{S}^2_+ - \hat{S}_-^2\big]\Big)\cdot\hat{S}_z},
\end{equation}
where $k=K_{xx}(\mathbf{r}-\mathbf{r}')=K_{yy}(\mathbf{r}-\mathbf{r}')=K_{zz}(\mathbf{r}-\mathbf{r}')$ is the diagonal element of the kernel Eq.~\eqref{eq:dipole-convolution-kernel}, and we've omitted the $\hat{S}_{\pm}$ terms due to zero $m_0$ component. The $\hat{S}^2_{\pm}$ terms mix the $m_{\pm}$ polarizations without hitting the eliminated $m_0$ channel implied by $\mathbf{M}_0\perp\mathbf{m}$ at linear order, producing elliptical precession. Such ellipticity is well known in ferromagnetic resonance and magnetostatic spin-wave physics (for example in thin films and surface/backward-volume waves) and should similarly emerge in confined dipole–exchange modes, potentially with a spatially nonuniform ellipticity profile \cite{Farle1998,kalinikos1986theory}.

\section{Conclusion}

In this work, we have investigated the rich spectrum of magnonic excitations in finite ferrimagnetic nanostructures, focusing on cylindrical and spherical geometries under various magnetization states. By systematically exploring the exchange-dominated, dipole–exchange, and dipolar regimes, we revealed the hierarchical structure of magnon modes, their classification by symmetry, and the emergence of anticrossings indicative of mode hybridization. Our numerical results demonstrated how eigenfrequencies scale with system size and quantum numbers, providing insight into the crossover from high-frequency exchange behavior to low-frequency magnetostatic dynamics.

A major outcome of this study is the development of a general coupled mode theory (CMT) for arbitrary three-dimensional classical magnonic systems. This framework follows in spirit the foundational ideas of Slavin and Kalinikos and allow treatment of confined geometries, enabling a clear understanding of how dipolar interactions selectively couple otherwise orthogonal exchange modes. The theory not only reproduces key features of observed spectra, such as anticrossings and symmetry-governed mode interactions, but also lays the groundwork for efficient semi-analytical modeling of complex magnonic devices.

Our analysis further highlights the importance of equilibrium magnetization textures in determining the magnonic spectra. In cases with transverse external fields or weak bias, the emergence of non-uniform, flower-like magnetization substantially alters the mode structure, demonstrating that uniform approximations may fail in realistic conditions. This insight is particularly relevant for small-scale or low-field magnonic systems, where magnetostatic effects dominate.

Looking forward, the generalized CMT framework offers a promising path for designing and interpreting experiments in confined magnonics. It can be extended to study nonlinear effects, thermal magnons, and coupling to other excitations such as phonons or photons. Moreover, the methods developed here are directly applicable to nontrivial geometries, such as ellipsoids, shells, or patterned structures, potentially enabling predictive modeling of engineered magnonic band structures and topological magnon phases. We expect this work to serve as a foundation for future theoretical and experimental efforts in nanoscale magnonics and spin-wave-based information processing.

\section{Acknowledgments}
Calculation and analysis of the magnonic spectra, analysis of the equilibrium magnetization states and limit cases, symmetry analysis and development of coupled-modes formalism was supported by Russian Scientific Foundation (project No. 24-72-00028).

\begin{appendix}
\section{Spin-1 operators}
\label{appx:spin-1-operators}
The vector product can be rewritten in the following operator form:
\begin{gather}
    \bm{c}=\bm{a}\times\bm{b} =-i(\widehat{\bm{S}}\cdot \bm{a})\bm{b} ,
\end{gather}
where $\bm{c},\bm{a},\bm{b}$ are arbitrary vectors, and $\widehat{\bm{S}}$ is the vector of 1-spin matrices. In index notation this expression becomes
\begin{equation}
    c_i=-\textit{i} \widehat{S}_{m_{,}in} a_m b_n
\end{equation}
Let us now consider a transformation to another basis.
If the components transform as $c'_i=U_{il} c_l$, then
\begin{gather}
    c'_i=-i U_{il}  \widehat{S}_{m_{,}ln} a_m b_n = \\\notag-i{U_{il}\widehat{S}_{m_{,}ln}(U^{-1})_{mf}(U^{-1})_{nj}} a'_f b'_j = -i \widehat{S}'_{f_{,}ij} a'_f b'_j
\end{gather}
From this expression we immediately obtain the transformation rule for the spin matrices:
\begin{gather}
   \widehat{S}'_{f_{,}ij}= U_{il}\widehat{S}_{m_{,}ln}(U^{-1})_{mf}(U^{-1})_{nj}= \\\notag
    =(U^{-T})_{fm} U_{il} \widehat{S}_{m_{,}ln} (U^{-1})_{nj}  
\end{gather}
In the Cartesian basis the spin-1 matrices take the form
\begin{widetext}
    \begin{equation}
    \label{eq:cart-spin-martices}
    \widehat{S}_x =\begin{pmatrix}
        0 & 0 & 0 \\ 0 & 0 &  -\textit{i} \\ 0 & \textit{i} & 0
    \end{pmatrix}, \ 
        \widehat{S}_y =\begin{pmatrix}
        0 & 0 & \textit{i}  \\ 0 & 0 &  0 \\ -\textit{i} & 0& 0
    \end{pmatrix}, \ 
    \widehat{S}_z =\begin{pmatrix}
        0 & -\textit{i} & 0  \\ \textit{i} & 0 &  0 \\ 0 & 0& 0
    \end{pmatrix},
\end{equation}
\end{widetext}
Under the transformation matrix given in Eq.~\eqref{eq:transform-cart-circ1}, in rotating basis the spin matrices become
\begin{widetext}
\begin{equation}
    \label{eq:hel-spin-martices}
    \widehat{\tilde{S}}_+ =\begin{pmatrix}
        0 & 1 & 0 \\ 0 & 0 &  1 \\ 0 & 0 & 0
    \end{pmatrix}, \ 
        \widehat{\tilde{S}}_z =\begin{pmatrix}
        -1 & 0 & 0  \\ 0 & 0 &  0 \\ 0 & 0& 1
    \end{pmatrix}, \ 
    \widehat{\tilde{S}}_- =\begin{pmatrix}
        0 & 0 & 0  \\ -1 & 0 &  0 \\ 0 & -1& 0
    \end{pmatrix},
\end{equation}
\end{widetext}
$\mathbf{\widehat{J}},\ \mathbf{\widehat{S}},\ \mathbf{\widehat{L}}$ satisfy the standard commutation relations
\begin{equation}
[\widehat{F}_m,\widehat{F}_j]=i \varepsilon_{mjn}\widehat{F}_n ,
\end{equation}
where $\widehat{F}_m$ stands for any of $\widehat{J}_m,\widehat{L}_m,\widehat{S}_m$ and $\varepsilon_{mjn}$ is the Levi-Civita symbol. Spin and orbital angular momentum commute:
\begin{equation}\label{eq:SL}
    [\widehat{S}_i,\widehat{L}_j] = 0
\end{equation}
Using the above relation, we obtain
\begin{gather}
    [\widehat{L}_i,\mathcal{H}_{\text{exch}}]=\notag\\
    =[\widehat{L}_i,\gamma\widehat{S}_z\cdot\left(\ell_{\text{exch}}^2 \ M_s(\Delta_{r}-\frac{\widehat{L}^2}{r^2}) - H^{\mathrm{eff}}_0\right)\mathbb{I}]=\notag\\
    \overset{Eq.~\ref{eq:SL}}{=}0
\end{gather}

\section{Rotational covariance of a dipole kernel}
\label{appx:rotatonal-covariance-proof}
Here we prove that the dipole interaction kernel 
\begin{equation}
    \label{eq:appendix-rcp-dipole-kernel-def}
    \widehat{K}(\Delta\mathbf{r}) = \frac{1}{4\pi}\left(\frac{\Delta\mathbf{r}\otimes\Delta\mathbf{r}}{ |\Delta\mathbf{r}|^5} - \frac{\mathbb{I}}{|\Delta\mathbf{r}|^3}\right), \ \ \ \ \Delta\mathbf{r} = \mathbf{r} - \mathbf{r}'
\end{equation}
is rotationaly covariant. Rotational covariance means that an action of continuous rotations about $\mathbf{\widehat{z}}$ axis $R_z(\alpha) = R$ on kernel such that
\begin{equation}
    \label{eq:appendix-rcp-rotational-covariance-def}
    \widehat{K}(R\Delta\mathbf{r}) = R\cdot\widehat{K}(\Delta\mathbf{r})\cdot R^T.
\end{equation}
To show that this is indeed the case, let us consider vomponents of the tensor \eqref{eq:appendix-rcp-dipole-kernel-def}
\begin{equation}
    \label{eq:appendix-rcp-dipole-kernel-component}
    K_{i,j} = \frac{1}{4\pi}\left(\frac{\Delta r_i\Delta r_j}{|\Delta\mathbf{r}|^5} - \frac{\delta_{ij}}{|\Delta\mathbf{r}|^3}\right)
\end{equation}
Considering that $(R\Delta \mathbf{r})_i=R_{ik}\Delta r_k$ and $|R\Delta\mathbf{r}| = |\Delta\mathbf{r}|$, 
\begin{widetext}
\begin{equation}
    \label{eq:appendix-rcp-kernel-covariance-prove}
    K_{ij}(R\Delta\mathbf{r}) = \frac{1}{4\pi}\left(\frac{R_{ik}\Delta r_k \ R_{jl}\Delta r_l}{|\Delta\mathbf{r}|^5} - \frac{\delta_{ij}}{|\Delta\mathbf{r}|^3}\right) = \frac{1}{4 \pi}R_{il}\left(\frac{\Delta r_k \Delta r_l}{|\Delta\mathbf{r}|^5} - \frac{\delta_{kl}}{|\Delta\mathbf{r}|^3}\right)R_{jl} = R_{ik}K_{kl}R_{lj}
\end{equation}
\end{widetext}

which proves \eqref{eq:appendix-rcp-rotational-covariance-def}.

\section{Appendix: Symmetry properties of the dipolar interaction}
\label{appx:dipolar-commutators}

In this Appendix we derive the commutation relations between the dipolar interaction operator $\mathcal{H}_{\rm{dip}}$ and the angular-momentum generators used in Sec.~III.C.

\subsection{Commutation with $\widehat J_z$}
\label{appx:commute-with-J}
A rotation about the equilibrium magnetization axis $\widehat{\mathbf z}$ acts on a vector field as
\begin{equation}
(U_\alpha\mathbf m)(\mathbf r)
=
R_z(\alpha)\,
\mathbf m\!\left(R_z^{-1}(\alpha)\mathbf r\right),
\qquad
U_\alpha=e^{-i\alpha\widehat J_z}.
\end{equation}
So, applying this transformation to a spin-wave under the convolution 
\begin{equation}
    \label{eq:appx:commute-with-U}
    \begin{split}
    &\big(\mathcal{H}_{\rm{dip}}U_{\alpha}\mathbf{m}\big)(\mathbf{r}) \\
    &\quad=\hat{S}_z\int_V{d^3\mathbf{r}' \ \hat{K}(\mathbf{r}-\mathbf{r}')\cdot\Big(R_z(\alpha) \mathbf{m}(R_z(\alpha)^{-1}\mathbf{r}')\Big)},
    \end{split}
\end{equation}
we re-denote $\mathbf{s}=R^{-1}_z(\alpha)\mathbf{r}'$; this does not affect the integration as both cylinder and sphere are symmetric under rotation, so integration domain does not change, and $\det R_z(\alpha) = 1$, thus integral in equation \eqref{eq:appx:commute-with-U} yields 
\begin{equation}
    \label{appx:commute-with-U}
    \begin{split}
   &\int_V{d^3\mathbf{r}' \ \hat{K}(\mathbf{r}-\mathbf{r}')\cdot\Big(R_z(\alpha) \mathbf{m}(R_z(\alpha)^{-1}\mathbf{r}')\Big)} =\\ &\int_V{d^3\mathbf{s} \ \hat{K}(\mathbf{r}-R_z(\alpha)\mathbf{s})\cdot R_z(\alpha) \mathbf{m}(\mathbf{s})} =\\
   &\int_V{d^3\mathbf{r}' \ \hat{K}\big(R_z(\alpha)(R_z^{-1}(\alpha)\mathbf{r}-\mathbf{s})\big)\cdot R_z(\alpha) \mathbf{m}(\mathbf{s})},
   \end{split}
\end{equation}
Using the covariance property \eqref{eq:appendix-rcp-rotational-covariance-def} and the axial invariance of the integration domain, one finds
\begin{equation}
    \begin{split}
        &\hat{S}_z\int_V{d^3\mathbf{r}' \ \hat{K}\big(R_z(\alpha)(R_z^{-1}(\alpha)\mathbf{r}-\mathbf{s})\big)\cdot R_z(\alpha) \mathbf{m}(\mathbf{s})} = \\
        &\hat{S}_z\int_V{d^3\mathbf{r}' \ R_z(\alpha)\hat{K}(R_z^{-1}(\alpha)\mathbf{r}-\mathbf{s})R_z^{-1}(\alpha)\cdot R_z(\alpha) \mathbf{m}(\mathbf{s})} =\\
        &U_{\alpha}\big(\mathcal{H}_{\rm{dip}}\mathbf{m}\big)(\mathbf{r}).
    \end{split}
\end{equation}
Therefore, 
\begin{equation}
\mathcal{H}_{\rm{dip}}\,U_\alpha = U_\alpha\,\mathcal{H}_{\rm{dip}} ,
\end{equation}
which implies
\begin{equation}
[\widehat J_z,\mathcal{H}_{\rm{dip}}]=0 .
\end{equation}
Thus the dipolar interaction preserves the azimuthal quantum number $j_z$.

\subsection{Failure to commute with $\widehat L^2$}
\label{appx:not-commute-with-L2}
Consider instead an orbital-only rotation
\begin{equation}
(\tilde U_\alpha\mathbf m)(\mathbf r)
=
\mathbf m\!\left(R_z^{-1}(\alpha)\mathbf r\right)
=
e^{-i\alpha\widehat L_z}\mathbf m(\mathbf r),
\end{equation}
which rotates the argument but leaves the vector components unchanged.
To show that, let us apply orbital rotation on a spin-wave inside a convolution 
\begin{equation}
    \label{eq:appx:orbital-rotation}
    \big(\mathcal{H}_{\rm{dip}}\tilde{U}_{\alpha}\mathbf{m}\big)(\mathbf{r}) = M_s\hat{S}_z\int_V{d^3\mathbf{r}' \ \hat{K}(\mathbf{r}-\mathbf{r}')\mathbf{m}(R^{-1}_z(\alpha)\mathbf{r}')}.
\end{equation}
Following the same as in \ref{appx:commute-with-J}, namely, subtituting $\mathbf{s}=R_z^{-1}(\alpha)\mathbf{r}'$ and employing rotational covariance of the kernel \eqref{eq:appendix-rcp-rotational-covariance-def}, one gets
\begin{equation}
\begin{split}
    &\big(\mathcal{H}_{\rm{dip}}\tilde{U}_{\alpha}\mathbf{m}\big)(\mathbf{r}) = \\ M_s\hat{S}_z\int_V{d^3\mathbf{r}' \ }&{  R_z(\alpha)\hat{K}(R_z^{-1}(\alpha)\mathbf{r}-\mathbf{r}')R_z^{-1}(\alpha)\mathbf{m}(\mathbf{s})}=\\ &\tilde{U}_{\alpha}\Big(R_z(\alpha)\mathcal{H}_{\rm{dip}} \ R^{-1}_z(\alpha)\mathbf{m}\Big)(\mathbf{r})
    \end{split}
\end{equation}
So, direct calculation shows that
\begin{equation}
\mathcal{H}_{\rm{dip}}\,\tilde U_\alpha \neq \tilde U_\alpha\,\mathcal{H}_{\rm{dip}} ,
\end{equation}
due to uncompensated rotations acting on the vector components.
Consequently,
\begin{equation}
[\widehat L_z,\mathcal{H}_{\rm{dip}}]\neq0,
\qquad
[\widehat L^2,\mathcal{H}_{\rm{dip}}]\neq0 .
\end{equation}
The dipolar interaction therefore breaks orbital $SO(3)$ symmetry and couples states with different orbital angular momentum $l$.

\subsection{Mixing of total angular momentum}

Since $\widehat J^2=\widehat L^2+\widehat S^2+2\widehat{\mathbf L}\cdot\widehat{\mathbf S}$ and $\mathcal{H}_{\rm{dip}}$ does not commute with $\widehat L^2$, it follows immediately that
\begin{equation}
[\widehat J^2,\mathcal{H}_{\rm{dip}}]\neq0 .
\end{equation}
Thus the dipolar interaction mixes states with different total angular momentum $\widehat{J}$, giving rise to hybridization and avoided crossings in the dipole--exchange spectrum.

\section{Axial uniformity of cylinder static demagnetizing field}
\label{axial-symmetry-proof}

Now, when kernel rotational covariance has been established, we use it to prove that the non-uniform static demagnetizing field  $\mathbf{H}^{\mathrm{dm}}_0(\mathbf{r})$ is axially symmetric, which formally expressed as
\begin{equation}
    \label{eq:appendix-axial-symmetry-def}
    \mathbf{H}^{\mathrm{dm}}_0(R\mathbf{r}) = R\cdot\mathbf{H}^{\mathrm{dm}}_0(\mathbf{r}).
\end{equation}
To do so, we apply coordinates rotation to integral convolution defining $\mathbf{H}^{\mathrm{dm}}_0$
\begin{equation}
    \label{eq:appendix-integral-convolution-dm-field-rotation}
    \mathbf{H}^{\mathrm{dm}}_0(R\mathbf{r}) = \int_V{d^3\mathbf{r}'\widehat{K}(R\mathbf{r} - \mathbf{r}')\cdot\mathbf{M}_0}.
\end{equation}
Let us now define $\mathbf{r}' = R\mathbf{s}$. Since $R$ is a popper rotation with $\det R = 1$ it trivially change integration variables $d^3\mathbf{r}' = d^3\mathbf{s}$. And due to cylinder admitting $SO(2)$ and the rotation under consideration is the one about $\mathbf{\widehat{z}}$ axis, this substitution does not change integration domain. Therefore, equation \eqref{eq:appendix-integral-convolution-dm-field-rotation} gives
\begin{widetext}
\begin{equation}
    \label{eq:appendix-axial-symmetry-proof}
    \begin{split}
    \int_V{d^3\mathbf{r}'\widehat{K}(R\mathbf{r}-\mathbf{r}')\cdot\mathbf{M}_0}
    &=\int_V{d^3\mathbf{s} \ \widehat{K}(R[\mathbf{r}-\mathbf{s}])\cdot\mathbf{M}_0} \\
    &=\int_V{d^3\mathbf{s}\left[R\cdot\widehat{K}(\mathbf{r}-\mathbf{s})\cdot R^T\right]\cdot\mathbf{M}_0} \\
    &=R\cdot\left[\int_V{d^3\mathbf{s}\widehat{K}(\mathbf{r}-\mathbf{s})\cdot R^T\cdot\mathbf{M}_0}\right],
    \end{split}
\end{equation}
\end{widetext}
and since the term on the very left is exactly $\mathbf{H}^{\mathrm{dm}}_0(R\mathbf{r})$  as defined by equation \eqref{eq:appendix-integral-convolution-dm-field-rotation} and the very right part is exactly $R\cdot\mathbf{H}^{\mathrm{dm}}_0(\mathbf{r})$ , this derivation formally proves that \eqref{eq:appendix-axial-symmetry-def} holds, thus proof of axial symmetry of static demagnetizing field of a cylinder is complete.

\section{Dipole Kernel symmetries under inversions}
\label{appx:inversion-symmetries}
Here we prove that the dipolar term remains invariant under full middle plane reflection,  but first show that the dipolar kernel \eqref{eq:dipole-convolution-kernel} is covariant under coordinate inversion, i.e. 
\begin{equation}
    \label{eq:appx:dipolar-kernel-inversion-covariance}
    \hat{K}(\hat{I}_i(\mathbf{r}-\mathbf{r}')) = \hat{I}_i \ \hat{K}(\mathbf{r}-\mathbf{r}')\hat{I}_i, \qquad i\in\{x,y,z\}, 
\end{equation}
where $\hat{I}_x = \text{diag}(-1,1,1)$. The proof follows the steps in Sec.~\ref{appx:rotatonal-covariance-proof} and employ the orthogonality of $\hat{I}_i$. We write the kernel explicitly 
\begin{equation}
    \hat{K} =\frac{1}{4\pi}\left(3\frac{\Delta\mathbf{r} \ [\Delta\mathbf{r}]^T}{\lvert\Delta\mathbf{r}\rvert^5} - \frac{\mathbb{I}}{\lvert\Delta\mathbf{r}\rvert^3}\right),
\end{equation}
where $\Delta\mathbf{r}=\mathbf{r}-\mathbf{r}'$. Using $\hat{I}_i^T=\hat{I}_i^{-1}=\hat{I}_i$, thus $\hat{I}_i^2=\mathbb{I}$, and $\lvert\hat{I}_i\Delta\mathbf{r}\rvert = \lvert\Delta\mathbf{r}\rvert$, we write
\begin{equation}
\begin{split}
    \frac{1}{4\pi}&\left(3\frac{\hat{I}_i\Delta\mathbf{r} \ [\hat{I}_i\Delta\mathbf{r}]^T}{\lvert\Delta\mathbf{r}\rvert^5} - \frac{\mathbb{I}}{\lvert\Delta\mathbf{r}\rvert^3}\right) = \\ \frac{1}{4\pi}&\left(3\frac{\hat{I}_i\Delta\mathbf{r} \ [\Delta\mathbf{r}]^T\hat{I}_i}{\lvert\Delta\mathbf{r}\rvert^5} - \frac{\hat{I}_i\mathbb{I}\hat{I}_i}{\lvert\Delta\mathbf{r}\rvert^3}\right) = \\
\frac{1}{4\pi}&\hat{I}_i\left(3\frac{\Delta\mathbf{r} \ [\Delta\mathbf{r}]^T}{\lvert\Delta\mathbf{r}\rvert^5} - \frac{\mathbb{I}}{\lvert\Delta\mathbf{r}\rvert^3}\right)\hat{I}_i,
    \end{split}
\end{equation}
therefore, relation \eqref{eq:appx:dipolar-kernel-inversion-covariance} has been derived.
 
 \subsection{Invariance of dipole term under  full middle plane reflection}

Once we've established dipole kernel coordinate inversion covariance \eqref{eq:appx:dipolar-kernel-inversion-covariance}, we proceed with proof of dipole interaction term invariance under full middle plane reflection, $\hat{P}_z$. That transforms spin-wave field components $\mathbf{m}$ as an axial vector, $\mathbf{m}=(m_x,m_y) \mapsto (\det \hat{I}_{z}) \hat{I}_{z}\mathbf{m} = (-m_x,-m_y)$, and position vector $\mathbf{r}\mapsto\hat{I}_z\mathbf{r}=(x,y,-z)$. Therefore, the action of $\hat{P}_z$ on $\mathbf{m}(\mathbf{r})$ can be written as
\begin{equation}
    \label{eq:appx:full-mirror}
    \hat{P}_z\mathbf{m}(\mathbf{r}) = -\hat{I}_{z}\mathbf{m}(\hat{I}_z\mathbf{r}).
\end{equation}
Acting with $\hat{P}_z$ on a spin-wave inside the convolution 
\begin{equation}
\label{eq:appx:mirror-in-convolution}
    \big(\mathcal{H}_{\rm{dip}}\hat{P}_z\mathbf{m}\big)(\mathbf{r}) = -\gamma M_s\hat{S}_z\int_V{d^3\mathbf{r}' \ \hat{K}(\mathbf{r}-\mathbf{r}'})\hat{I}_{z}\mathbf{m}(\hat{I}_z\mathbf{r}').
\end{equation}
With substitution $\mathbf{s}=\hat{I}_z\mathbf{r}'$, which does not affect the integral due to cylinder being coordinate inversion invariant, and change of integration variable leads to $d^3\mathbf{r}'\mapsto \lvert\det \hat{I}_z\rvert d^3\mathbf{s} = d^3\mathbf{r}$, we can rewrite the Eq.~\eqref{eq:appx:mirror-in-convolution}
\begin{equation}
\begin{split}
   -\gamma M_s\hat{S}_z&\int_V{d^3\mathbf{s} \ \hat{K}(\mathbf{r}-\hat{I}_z\mathbf{s}})\hat{I}_{z}\mathbf{m}(\mathbf{s})=\\&-\gamma M_s\hat{S}_z\int_V{d^3\mathbf{s} \ \hat{K}(\hat{I}_z(\hat{I}_z\mathbf{r}-\mathbf{s}}))\hat{I}_{z}\mathbf{m}(\mathbf{s}).
    \end{split}
\end{equation}
Using covariance relation \eqref{eq:appx:dipolar-kernel-inversion-covariance}
\begin{equation}
    -\gamma M_s\hat{S}_z\int_V{d^3\mathbf{s} \ \hat{I}_{z}\hat{K}(\hat{I}_z\mathbf{r}-\mathbf{s}})\hat{I}_{z}^2\mathbf{m}(\mathbf{s})=\hat{P}_z \big(\mathcal{H}_{\rm{dip}}\mathbf{m}\big)(\mathbf{r}),
\end{equation}
therefore, 
\begin{equation}
    \hat{P}_z\mathcal{H}_{\rm{dip}} = \mathcal{H}_{\rm{dip}}\hat{P}_z.
\end{equation}
\subsection{Non-invariance of dipole interaction term under coordinate inversion}
$\mathcal{I}_i, i\in\{x,y,z\}$ it is coordinate inversion operator, such that $\mathcal{I}_i\mathbf{m}(\mathbf{r}) = \mathbf{m}(\hat{I}_i\mathbf{r})$. Applying this transform to convolution, one gets
\begin{equation}
\label{eq:appx:inversion-in-convolution}
    \big(\mathcal{H}_{\rm{dip}}\mathcal{I}_i\mathbf{m}\big)(\mathbf{r}) = \gamma M_s\hat{S}_z\int_V{d^3\mathbf{r}' \ \hat{K}(\mathbf{r}-\mathbf{r}'})\mathbf{m}(\hat{I}_i\mathbf{r}').
\end{equation}
We again substitute $\mathbf{s}=\hat{I}_i\mathbf{r}'$, yielding
\begin{equation}
\begin{split}
    \gamma M_s\hat{S}_z&\int_V{d^3\mathbf{s} \ \hat{K}(\mathbf{r}-\hat{I}_i\mathbf{s}})\mathbf{m}(\mathbf{s})=\\&\gamma M_s\hat{S}_z\int_V{d^3\mathbf{s} \ \hat{K}(\hat{I}_i(\hat{I}_i\mathbf{r}-\mathbf{s}}))\mathbf{m}(\mathbf{s}).
    \end{split}
\end{equation}
Using covariance relation \eqref{eq:appx:dipolar-kernel-inversion-covariance}
\begin{equation}
    \gamma M_s\hat{S}_z\int_V{d^3\mathbf{s} \ \hat{I}_{i}\hat{K}(\hat{I}_i\mathbf{r}-\mathbf{s}})\hat{I}_{i}\mathbf{m}(\mathbf{s})=\mathcal{I}_i \big(\hat{I}_i\mathcal{H}_{\rm{dip}}\hat{I_i}\mathbf{m}\big)(\mathbf{r}),
\end{equation}
thus dipole interaction operator and spin-wave field acquire additional mirror operations under coordinate-only inversion, meaning 
\begin{equation}
    \mathcal{I}_i\mathcal{H}_{\rm{dip}} \ne \mathcal{H}_{\rm{dip}}\mathcal{I}_i
\end{equation}

{\section{Stability of the exchange-mode CMT}
\label{app:cmt-stability}

The coupled-mode construction in Sec.~\ref{sec:cmt} represents the dipole--exchange eigenstates as superpositions of exchange modes. 
Its accuracy therefore depends on the stability of the low-frequency spectrum as this basis is enlarged. 
We examine this stability independently in the three representative symmetry sectors $(J_z,P_z)=(0,+1)$, $(1,-1)$, and $(1,+1)$ for both spherical and cylindrical resonators. 
In each sector, the positive-frequency exchange modes are added in order of increasing exchange frequency. 
The comparison follows the five lowest positive eigenlines at nine logarithmically spaced radii between $25$ and $600$~nm, with truncations up to $N=40$.

\subsection{Local convergence}

Let $\omega_k^{(N)}(R_i)$ denote the $k$th selected eigenfrequency obtained with $N$ exchange modes at radius $R_i$. The change produced by adding the $N$th mode is measured by the relative root-mean-square quantity
\begin{equation}
 \varepsilon_{\mathrm{loc}}(N)=
 \left[
 \frac{\sum_{i,k} w_i
  \left|\omega_k^{(N)}(R_i)-\omega_k^{(N-1)}(R_i)\right|^2}
 {\sum_{i,k}w_i\left|\omega_k^{(N)}(R_i)\right|^2}
 \right]^{1/2},
\label{eq:cmt-local-stability}
\end{equation}
where $w_i$ are normalized trapezoidal weights on the sampled radius grid. 
This local measure resolves the spectral renormalization caused by one additional exchange mode. 
It is therefore a sensitive test of whether a specific omitted mode still couples appreciably to the low-lying branches in the same symmetry sector.

The dipolar contribution is not a scalar series with terms of fixed sign.
Each newly admitted mode introduces further diagonal and off-diagonal matrix elements and can shift the selected branches through hybridization. 
The local changes consequently need not decrease monotonically: a mode with a large dipolar overlap, or one lying relatively close in frequency to a selected branch, may produce a larger correction than the modes preceding it. 
Retaining the individual increments in Fig.~\ref{fig:cmt-local-convergence} makes this physically relevant structure visible and provides a stringent local stopping criterion.

\begin{figure}[t]
 \includegraphics[width=\columnwidth]{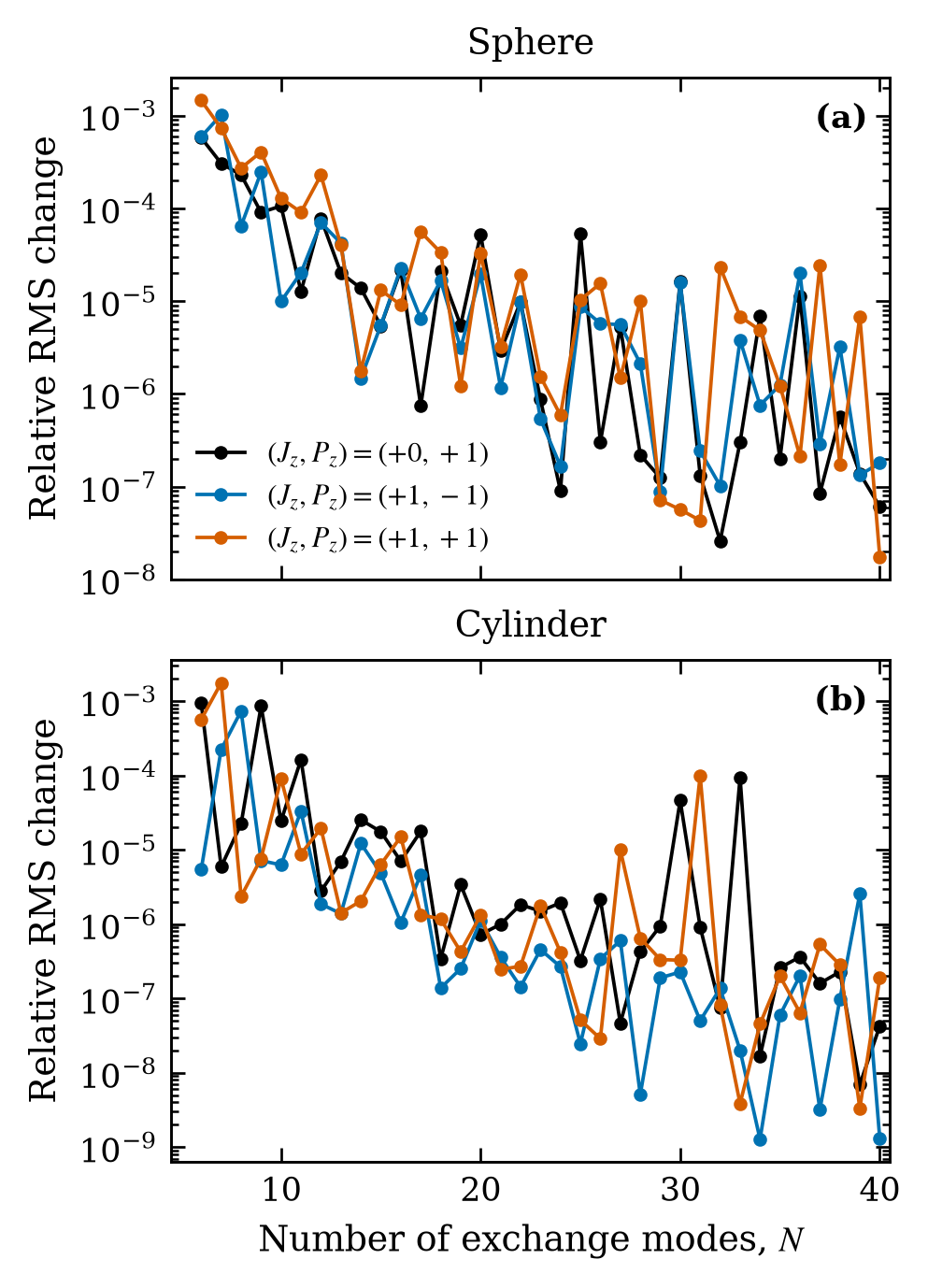}
 \caption{Local convergence of the exchange-mode CMT. Solid curves with
 markers show the relative RMS change in Eq.~\eqref{eq:cmt-local-stability}
 when the basis is enlarged from $N-1$ to $N$. (a) Sphere and (b) cylinder,
 for the three indicated symmetry sectors. The individual increments are
 non-monotone because a newly admitted exchange mode changes both diagonal and
 off-diagonal entries of the truncated Hamiltonian. At $N=40$, the local
 relative RMS changes range from $1.3\times10^{-9}$ to
 $1.9\times10^{-7}$.}
 \label{fig:cmt-local-convergence}
\end{figure}

\subsection{Cumulative convergence}

A small local increment at a particular $N$ does not by itself establish that the spectrum is stable against all subsequent modes. We therefore also compare each truncation with the terminal basis $N_{\mathrm{ref}}=40$ and define
\begin{equation}
 \varepsilon_{\mathrm{cum}}(N;40)=
 \left[
 \frac{\sum_{i,k}w_i
  \left|\omega_k^{(N)}(R_i)-\omega_k^{(40)}(R_i)\right|^2}
 {\sum_{i,k}w_i\left|\omega_k^{(40)}(R_i)\right|^2}
 \right]^{1/2}.
\label{eq:cmt-cumulative-stability}
\end{equation}
This cumulative measure gives the total displacement of the selected eigenlines from the reference spectrum after all modes between $N+1$ and $40$ have been included. 
It complements Eq.~\eqref{eq:cmt-local-stability} by revealing the net effect of successive corrections, including partial cancellations, and is thus the more direct measure of spectral stability over a prescribed basis range. 
The $N=40$ spectrum is used here as a controlled finite-basis reference, rather than as an assertion of the infinite-basis limit.

\begin{figure}[t]
 \includegraphics[width=\columnwidth]{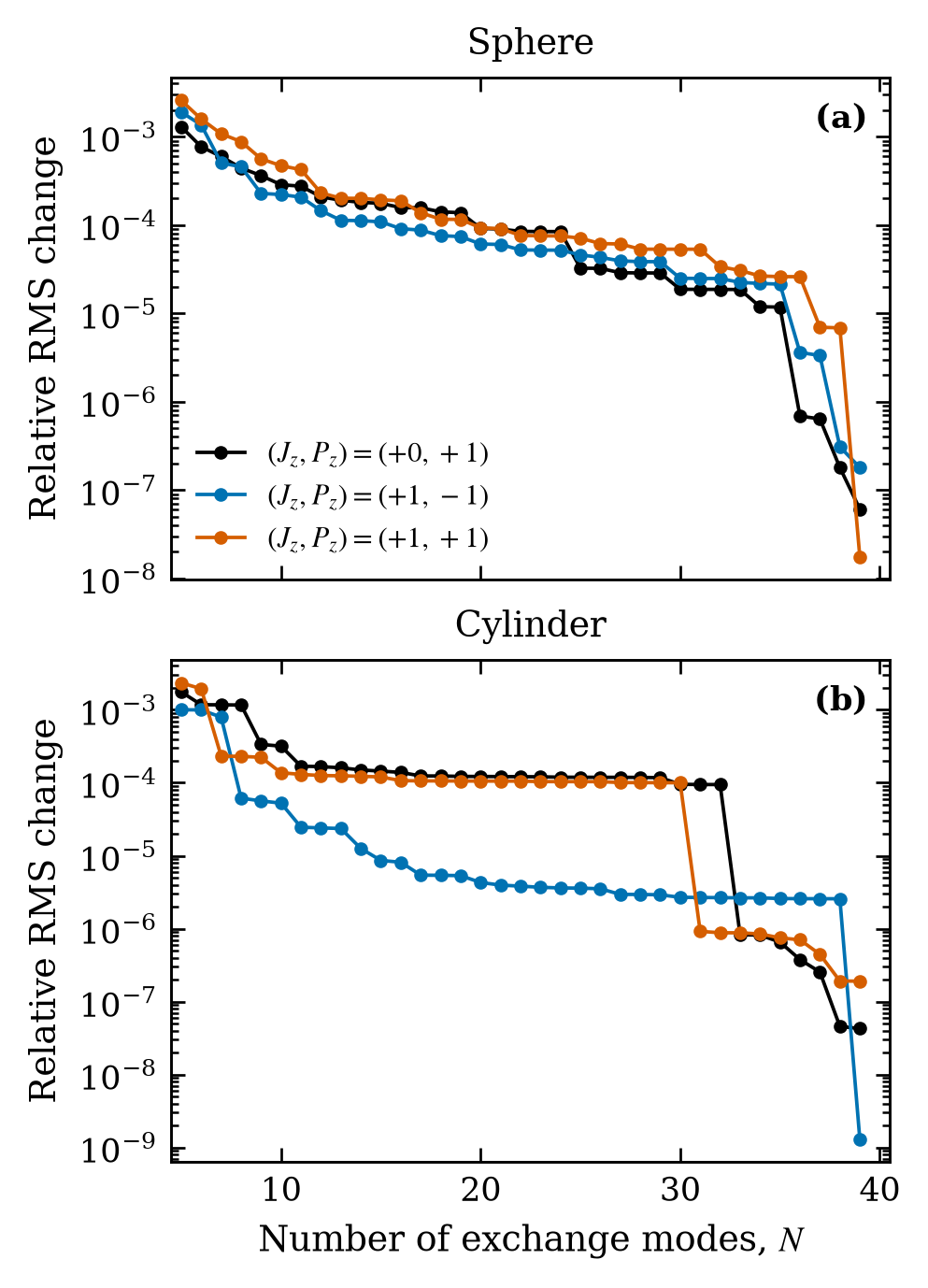}
 \caption{Cumulative convergence of the exchange-mode CMT relative to the
 $N_{\mathrm{ref}}=40$ reference, using Eq.~\eqref{eq:cmt-cumulative-stability}.
 Solid curves show the direct residual. (a) Sphere and (b) cylinder. The
 cumulative diagnostic confirms the decay toward the reference spectrum while
 retaining the sector-dependent plateaus and drops. The $N=40$ endpoint is
 identically zero and is omitted from the logarithmic plot.}
 \label{fig:cmt-cumulative-convergence}
\end{figure}

The local and cumulative curves together show convergence of the low-lying CMT spectrum over the range considered, with sector-dependent and generally non-monotone corrections. 
This non-monotonicity concerns the sequence of nested truncations, not the enumeration of a fixed basis. 
A permutation of the same $N$ exchange modes transforms $H_N$ by similarity and leaves its eigenfrequencies unchanged. 
The chosen ordering specifies which physical modes enter each intermediate truncation and therefore provides an unambiguous summation prescription.

Some care is required when the CMT expansion is extended to substantially larger $N$. 
The exchange-frequency scale spanned by the basis increases with the truncation, and the condition number of the resulting matrix may grow.
Small inaccuracies in the dipolar matrix elements can then produce an enhanced shift of the calculated eigenfrequencies. 
At the same time, higher exchange modes have increasingly rapid spatial oscillations, so their nonlocal dipolar overlaps require a correspondingly finer radial, axial, and angular quadrature. 
An apparent saturation or increase of either convergence measure at large $N$ may therefore reflect insufficient resolution of these overlaps rather than a physical lack of convergence of the mode expansion.

Accordingly, extensions beyond the present $N\leq40$ range should be accompanied by a joint convergence test in the basis size and in the quadrature.
At fixed $N$, the dipolar overlaps and eigenfrequencies should first be stable under quadrature refinement; only then can their variation with $N$ be assigned to the CMT truncation itself. 
The present test establishes the stability of the reported CMT spectra over the stated range and complements the direct CMT--multiphysics comparison in Fig.~4.
}
\end{appendix}

\bibliography{references}
\end{document}